\documentclass{aa}  
\usepackage{graphicx}
\usepackage{txfonts}
\usepackage{multicol}
\usepackage{hyperref}
\usepackage{array}
\usepackage{placeins}
\usepackage{tablefootnote}
\newcolumntype{P}[1]{>{\centering\arraybackslash}p{#1}}
\begin{document}

   \title{Constraining the overcontact phase in massive binary evolution}
   \subtitle{III. Period stability of known B+B and O+B overcontact systems}
   \author{Jasmine Vrancken \inst{1,2} \and
            Michael Abdul-Masih \inst{3,4,1} \and
            Ana Escorza \inst{3,4,1} \and
            Athira Menon \inst{3,4} \and
            Laurent Mahy \inst{5} \and
            Pablo Marchant \inst{2}
          }
   \institute{European Southern Observatory, Alonso de Cordova 3107, Vitacura, Casilla 19001, Santiago de Chile, Chile
   \and Institute of Astronomy, KU Leuven, Celestijnenlaan 200 D, 3001 Leuven, Belgium \\ \email{jasmine.vrancken@kuleuven.be} 
    \and
    Instituto de Astrofísica de Canarias, C. Vía Láctea, s/n, 38205 La Laguna, Santa Cruz de Tenerife, Spain.
    \and
    Universidad de La Laguna, Departamento de Astrofísica, Av. Astrofísico Francisco Sánchez s/n, 38206 La Laguna, Tenerife, Spain.
    \and
    Royal Observatory of Belgium, Avenue Circulaire/ Ringlaan 3, 1180 Brussels, Belgium
    }
   \date{Received month day, year; accepted month day, year}

\abstract
   {Binary systems play a crucial role in massive star evolution. Systems composed of B-type and O-type stars are of particular interest due to their potential to lead to very energetic phenomena or the merging of exotic compact objects.}
   {We aim to determine the orbital period variations of a sample of B+B and O+B massive overcontact binaries, with the primary objectives of characterizing the evolutionary timescales of these systems and addressing the existing discrepancy between observational data and theoretical predictions derived from population synthesis models.}
   {We used \textsc{Period04} to analyze archival photometric data going back a century for a sample of seven binary systems to measure their orbital periods. We then determine the period variations using a linear fit.}
   {We find that the period variation timescales of five truly overcontact binary systems align with the nuclear timescale, in agreement with previous findings for more massive overcontact binaries. Additionally, we noticed a clear distinction between the five systems that had been unambiguously classified as overcontact systems and both SV~Cen and VFTS~066, which seem to be evolving on thermal timescales and might be misclassified as overcontact systems.}
   {In the case of the five overcontact binaries, our results indicate a noticeable mismatch between the observational data and the theoretical predictions derived from population synthesis models. Furthermore, our results suggest that additional physical mechanisms must be investigated to compare the observed variations more thoroughly with theoretical predictions. 
   }


\maketitle

\section{Introduction}
Massive stars have a strong impact on their environment due to their high luminosity and their intense radiative feedback, which drives both the chemical and mechanical evolution of their host galaxies. In addition, their high luminosity allows for the discovery of additional information about distant objects, contributing to an enhanced understanding of galactic and extragalactic astronomy \citep[e.g.,][]{herrero, grudic}. These feedback mechanisms primarily manifest themselves in the form of protostellar outflows, strong stellar winds, ionizing radiation, and supernova explosions \citep{mcleod}. 
After they explode as supernovae, massive stars can end their lives as neutron stars or black holes \citep{woosley, heger, maeda, heger2023}, but the different evolutionary paths that lead them to this stage contain many uncertainties.
\newline \indent
Binarity is a common phenomenon at all stellar masses and must be properly accounted for when considering the future evolution of a given system \citep[e.g.,][]{paczynski, annualrevpablo}. This is even more important in the case of massive stars, since more than 70\% of them will interact with a companion before leaving the main sequence \citep{sana, Moe2017}. 
Massive stars are interesting due to their sensitivity to factors such as metallicity, rotation, and possibly magnetic fields \citep{2012ARA&A..50..107L}. \citet{RevModPhys.84.25} provided a detailed discussion on the evolution of rotating stars, including the rotational mixing in O- and B-type stars.
Many aspects of binarity have been intensively studied from both a theoretical and observational perspective, including the effects of tides \citep{hut, zahn, hurley}, the outward transfer of angular momentum to form disks \citep{Lee1991}, mass and energy outflow \citep{shufh}, and mass transfer driven by stellar winds \citep{Mohamed2007}, but many uncertainties and unknowns remain. More recently, the Modules for Experiments in Stellar Astrophysics (MESA) code \citep{mesa1, mesa2} was extended to include binaries \citep{mesa3} in order to experiment with these binary interaction mechanisms and improve our theoretical knowledge about binarity.
%
\newline \indent
One particularly important phase in the evolution of massive binary systems is the overcontact phase, when both components of a binary system are overfilling their Roche lobes. This phase is often also called the contact phase. 
This evolutionary phase is characterized by many simultaneous complex physical processes, including mass and energy exchange, mutual irradiation, tidal and Roche deformations, intense stellar winds, and a high degree of internal mixing. Our understanding of this phase is further complicated by a lack of observational constraints for many of these physical processes. Only about 20 massive ($M_{init} > 8 M_\odot$) overcontact binaries are currently known despite the fact that 25\% of all massive binaries are expected to go through such a phase at some point in their evolution \citep{1994A&A...290..119P, 2001A&A...369..939W, Michael2021A&A...651A..96A, 2024A&A...682A.169H}.
This is partly due to the fact that distinguishing overcontact systems from semidetached and detached systems is not trivial \citep{2020AA...634A.119M}. It has often been attempted by analyzing only their light curves \citep{wilson}. In certain cases, if we find that the system is eccentric or if the rotation rates of the components are highly asynchronous, we can discard the possibility that the system is in an overcontact system. However, the only way to undoubtedly discern between the different morphologies is combining the analysis of their light curves with spectroscopic data \citep{2020A&A...636A..59A}. 
\newline \indent
One method that can help us better understand the physical processes that govern this specific stage of binary evolution is to study how the orbital periods of massive binaries evolve during the overcontact phase. Analyses of the rate at which a period changes have proven useful in understanding eclipsing pulsar binaries \citep{1994ApJ...436..312A}, post-common-envelope binaries \citep{2010MNRAS.407.2362P}, and O+O massive overcontact binaries \citep{Michael2022A&A...666A..18A} revealed a discrepancy between our theoretical understanding of how one thinks these systems should evolve and what the observations actually show. 
Specifically, the observed mass ratios between the secondary and the primary components do not asymptotically approach $q = 1$ as population synthesis simulations predict, but are instead evenly distributed between $q= 0.6$ and $q=1$ \citep{2021MNRAS.507.5013M}. Furthermore, from the O+O massive binaries sample, \citet{Michael2022A&A...666A..18A} conclude that exploring this possible discrepancy requires an increase in the sample size of studied systems.
\newline \indent
While B-type stars are more numerous than O-type stars, so far O-type overcontact systems have been studied more extensively because their high luminosities make them more easily observable than their B-type counterparts. Despite this, B-type overcontact systems can provide many useful insights and can act as a complementary sample to the O-type systems studied by \citet{Michael2022A&A...666A..18A}. In expanding our focus to include both B+B and O+B overcontact systems, we aim to provide more observational constraints for future theoretical efforts and to investigate any similarities or differences between the two sets of samples. This investigation thus allows us to expand the sample of studied massive overcontact binaries, which is needed to confirm the reported discrepancy.
\newline \indent
In Sect. \ref{sec:sample} we discuss our sample selection and the corresponding available archival photometric data. In Sect. \ref{sec:methods} we discuss how we determine the orbital period from the photometric data and which fitting model we used to determine the change in period. In Sect. \ref{sec:results} we present our results, and in Sect. \ref{sec:discuss} we discuss what we can learn from them. Lastly, in Sect. \ref{sec:concl} we summarize our research and consider possible further studies.

\section{Sample and archival data}
\label{sec:sample}

\begin{table*}[ht]
\caption{Coordinates (J2000) and orbital periods in units of days and seconds for our sample of B+B and O+B overcontact binaries and their mass ratios $q=M_2/M_1$.}
\label{table:literaturevalues}
    \centering
    \begin{tabular}{c c c c c c c}
        \hline \hline
         Identifier & RA (J2000)  & Dec (J2000) & $P$ [days] & $P$ [seconds] & $q$ & Reference \\
         \hline 
         CT~Tau & 05 58 50.11 & 27 04 41.92 & 0.66682928(16) & 57614.049792 & 0.983 $\pm$ 0.003 & \cite{Yang2019AJ....157..111Y} \\
         GU~Mon & 06 44 46.86 & 00 13 18.30 & 0.89664680(56) & 77470.283520 & 0.976 $\pm$ 0.003 & \cite{Yang2019AJ....157..111Y} \\
         SV~Cen & 11 47 57.22 & -60 33 57.76 & 1.658 & 143251.200000 & 0.71 $\pm$ 0.02 & \cite{1992AJ....103..573R} \\
         V606~Cen & 13 21 36.28 & -60 31 14.75 & 1.4950935 & 129176.078400 & 0.5484 $\pm$ 0.0007 & \cite{Li2022ApJ...924...30L} \\
         V701~Sco & 17 34 24.51 & -32 30 15.99 & 0.76187385(13) & 65825.900640 & 0.995 $\pm$ 0.002 & \cite{Yang2019AJ....157..111Y} \\
         V745~Cas & 00 22 53.34 & 62 14 28.98 & 1.4106019 & 121876.004160 & 0.57 $\pm$ 0.02 & \cite{10.1093/mnras/stu958} \\
         VFTS~066 & 05 37 33.09 & -69 04 34.69 & 1.141160 & 98596.224000 & 0.523 $\pm$ 0.014 & \cite{2020AA...634A.119M} \\ 
         \hline
    \end{tabular}
    \tablefoot{Errors are included where available.}
\end{table*}

\begin{figure*}
    \centering
    \includegraphics[width=\textwidth]{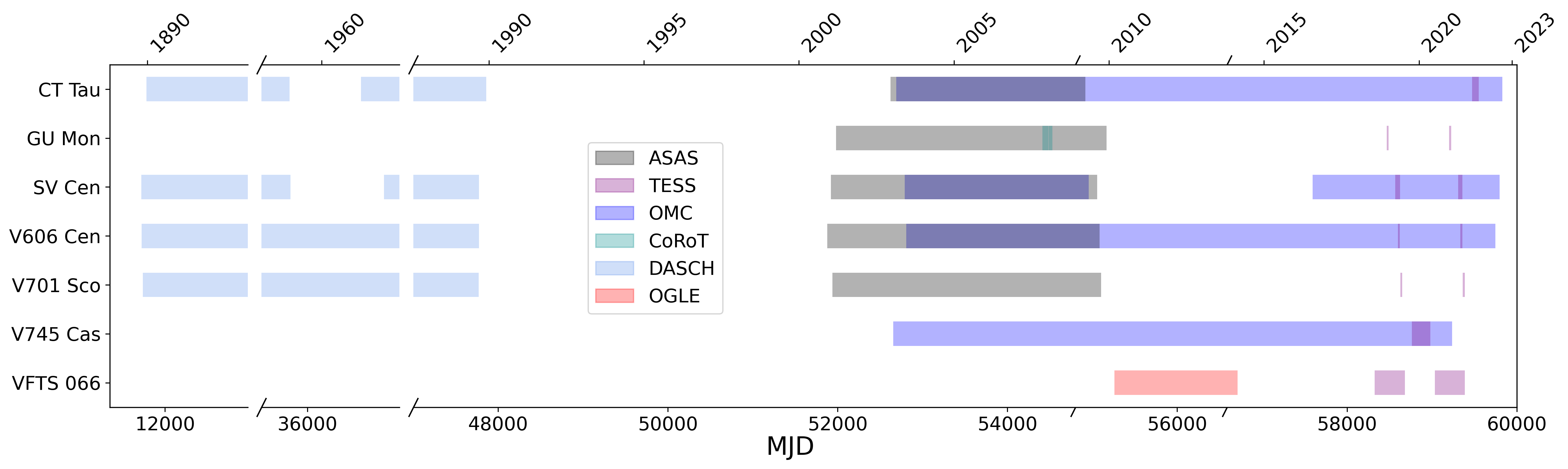}
    \caption{Time span overview of archival photometric data used for each target. Different instrument observations are color-coded.}
    \label{fig:timebases}
\end{figure*}

To gather a representative sample of systems, we searched in the literature (see Table \ref{table:literaturevalues}) for reported O+B and B+B overcontact binary systems that met the following conditions. 
First, the system needs to be classified as an overcontact binary based on combined photometric and radial velocity fits. OGLE SMC-ECL-2063 was considered an overcontact binary by \cite{Wu10.1093/pasj/psad003}, but up to now there was no spectroscopic data available to determine the radial velocities. Therefore, this system was not included in our sample. 
Secondly, the spectral type of at least one of the components must be B and the mass of the primary must be at least 8$M_\odot$ to be qualified for our sample. 
It was also important that the photometric signal was dominated by the orbital period of the binary system and that there were no significant influences from other periodic signals or the presence of a nearby third object. 
With these restrictions in mind, our sample consists of seven overcontact binary systems.
\newline \indent
Most of the systems we encountered are actually reported to be in a higher order multiple star system. CT~Tau, GU~Mon, V606~Cen, and V701~Sco were thought to have a third companion \citep{Yang2019AJ....157..111Y, Wu10.1093/pasj/psad003} and V745~Cas is a higher order multiple system with a binary in overcontact configuration orbiting the other components in a long period orbit \citep{10.1093/mnras/stu958}. 
The orbital periods of the outer companions were long enough (i.e., $> 30$ years) that their impact on the orbital parameters of the overcontact binaries can be neglected on timescales probed in the current study. 
A summary of our sample and the most recently derived orbital periods can be found in Table \ref{table:literaturevalues}. A brief summary of each object in our sample can be found below and an overview of the time span of the photometric data is shown in Fig. \ref{fig:timebases}.

\subsection{CT~Tau}
The first observations of the binary system CT~Tau were done by \citet{1965IBVS...95....1I}. Later on \citet{Plewa1993} analyzed all the available photometric data up to then to obtain absolute parameters of the system. They also concluded that the orbital period did not change in 30 years and therefore CT~Tau should be in an equilibrium state of evolution without any trace of dynamical instability or mass transfer. 
These results were reviewed by \citet{Yang2019AJ....157..111Y} by including their own multicolor photometric observations. The data that we used to analyze the system consisted of photometry from the All Sky Automated Survey (ASAS) Catalog of Variable Stars \citep{asas}, Digital Access to a Sky Century at Harvard \citep[DASCH;][]{dasch}, Optical Monitoring Camera (OMC) Archive \citep{omc} and Transiting Exoplanet Survey Satellite \citep[TESS;][]{TESS}. The TESS observations were acquired in the fourth year in sectors 43, 44 and 45. 

\subsection{GU~Mon}
The parameters of GU~Mon were determined with an analysis of the B and V light curves and high-resolution spectra by \cite{Lorenzo2016}. The orbital period was also verified by comparing it with the known radial velocities of other members in the Dolidze 25 cluster \citep{Lorenzo2016}. As for CT~Tau, \citet{Yang2019AJ....157..111Y} included new multicolor photometric observations to study GU~Mon more in detail. They measured an orbital period change that was up to then still negligible for GU~Mon. To assemble a light curve spread over a wide range of time, the photometric data from ASAS Catalog of Variable Stars \citep{asas}, CoRoT \citep{CoRoT2009AJ....138..649D} and TESS \citep{TESS} were combined. The available TESS data originated from sector 6 and 33. 


\subsection{SV~Cen}
Already in 1972 the eclipsing binary SV~Cen was observed and classified as a double-lined spectroscopic binary \citep{1972PASP...84..686I}. In 1982 \citet{1982Ap&SS..83..163N} determined the observed rate of mass transfer based on previous observations. The period-time diagram was extended laboriously by \citet{1982A&A...110..246D, 1985Ap&SS.114....1H, 1992AJ....103..573R, 1993IBVS.3868....1D, 1994AGAb...10...95D} and in the last-mentioned paper SV~Cen was described as the eclipsing pair with the largest known rate of period decrease. However, the literature was ambiguous about the configuration of the system. Some claimed that it is an overcontact system \citep{1994iue..prop.4773D} and others claimed that it is semi-detached \citep{1994A&A...291..786D, deschamps, davis}. The combined light curve that we use here consists of photometry from ASAS Catalog of Variable Stars \citep{asas}, DASCH \citep{dasch}, OMC Archive \citep{omc} and TESS \citep{TESS}. The available TESS data originated from sectors 10, 11, 37, 38 and 64. 

\subsection{V606~Cen}
V606~Cen was photometrically analyzed for the first time by \citet{1994A&AS..104....9F}. Five years later, the early-type eclipsing binary V606~Cen was simultaneously analyzed with spectroscopic and photometric data \citep{1999A&A...345..531L}. 
Thereafter, the system was not studied by other authors until \citet{Li2022ApJ...924...30L}. They determined the period decreasing rate and found that V606~Cen has a very low fill-out factor of about 2\%. The photometric data available spanned over a period from 1889 until 2022. This was composed of data from ASAS Catalog of Variable Stars \citep{asas}, DASCH \citep{dasch}, OMC Archive \citep{omc} and TESS \citep{TESS}. The available TESS data was observed in sector 11, 38 and 65. 

\subsection{V701~Sco}
The system V701~Sco was mentioned in multiple studies and was characterized as a variable star by \citet{1948AnLei..20....3P}. By measuring the color, \citet{1961RGOB...27...61E} was able to locate the system in the galactic cluster NGC 6383. \citet{1974A&AS...13..315L} observed the binary with B- and V-filters to gain a complete coverage of the light curve. Later on, the radial velocity and the absolute parameters were determined \citep{1980A&A....82..225A, 1987MNRAS.226..899B}. V701~Sco was studied in more detail as an early-type overcontact twin binary by \citet{Yang2019AJ....157..111Y}. Our analyzed photometric data consisted of ASAS Catalog of Variable Stars \citep{asas}, DASCH \citep{dasch} and TESS \citep{TESS}. For this particular system, sectors 12 and 39 from TESS were used.

\subsection{V745~Cas}
V745~Cas was observed for the first time with Hipparcos \citep{1997A&A...323L..49P, 2009A&A...500..505V}. Afterward, the Hipparcos and International Gamma-Ray Astrophysics Laboratory (INTEGRAL) light curves were analyzed and using optical spectroscopic observations the radial velocities were determined by \citet{10.1093/mnras/stu958}. \citet{2015AstL...41..473B} used the known properties of V745~Cas to determine the galactic rotation curve. For this project we used the photometric data from the OMC Archive \citep{omc} and TESS \citep{TESS} in particular the sectors 17, 18, 24 and 58. 

\subsection{VFTS~066}
Within our sample, VFTS~066 uniquely represents the category of O+B overcontact systems that satisfies our selection criteria. Initially identified through observations conducted as part of the Very Large Telescope Fibre Large Array Multi Element Spectrograph (VLT-FLAMES) Tarantula Survey, this binary system was confirmed as an overcontact binary in the work of \citet{2020AA...634A.119M}. Subsequently, an atmospheric analysis of the system was carried out in a follow-up study \citep{2020A&A...634A.118M}. Available data of the VFTS~066 binary system came from the Optical Gravitational Lensing Experiment (OGLE) database and TESS sectors in year 1, 3 and 5. However, the TESS data from the fifth year proved too noisy to extract meaningful signals, rendering it unusable for our purposes. Additionally, Sector 31 of TESS Year 3 data was excluded for similar reasons. Consequently, we used data exclusively from the OGLE database and the first and third year of TESS observations, with the exclusion of Sector 31. 

\subsection{Data reduction}
The photometric data listed above for each target was used to determine the orbital period of the binary system. To obtain consistent results, outliers and points with bad quality flags were removed. When we retrieved data from ASAS \citep{asas, 2003AcA....53..341P} only data points with grade A and grade B were selected. From the OGLE database \citep{szymanski2006optical}, we used the I band and removed obvious outliers. The data from the OMC Archive \citep{omc} had different labels. Only data points with label 'Good' were included, and data points with labels 'Centroid too far from source coord.', 'Brightest pixel forced', 'Bad Centroid', 'Bad PSF' (Point Spread Function), 'Bad Pixels', 'Bad Background' and 'Mosaic' were excluded. Additionally, we also removed obvious outliers from the data sets with a 'Good' flag. From the DASCH catalog \citep{dasch}, only obvious outliers were removed. Concerning TESS \citep{TESS}, only one of our targets had a light curve reduced with \textsc{SPOC} \citep{spoc}, the official TESS pipeline. For all the other targets and sectors, we used the python package \textsc{Lightkurve} \citep{2020AAS...23540904B}. We started from a 9x9 frame of pixels around the observed object. To minimize the chance of contamination, we took only 1 pixel as the target mask for each individual target. We always took the pixel with the highest brightness around the center pixel. After this selection, we could extract the light curve using the package. Thereafter, we removed the outliers and the NaN points and flattened the light curve with the built-in \texttt{flatten()} function. With the TESS data we also paid attention to the deviations caused by the downlink in the beginning, middle, and end of the TESS sectors. These deviating points were subsequently removed from the data set.

\section{Methods}
\label{sec:methods}
\subsection{Period determination}
The data sets from DASCH spanned a time period up to around 100 years and INTEGRAL OMC covered up to two decades. When the data sets were so long, we split them in multiple smaller sets in order to obtain several period measurements over such long times. The OMC data was always separated in two sets with an equal amount of data points, and the DASCH light curves were split in three or more subsets, with a maximum of ten subsets for CT~Tau. This choice was made depending on the size of the error bar that we obtained for the period. TESS data was organized by grouping sectors according to their respective observation years. 
\newline \indent
Once we had all the required photometry from our sample, the goal was to determine the change in orbital period over the time base of observation. To achieve this goal, we used the publicly available \textsc{Period04} \citep{2005CoAst.146...53L} software.
Another effective tool to detect signals in unevenly sampled time series is the Lomb-Scargle periodogram \citep{1976Ap&SS..39..447L, 1982ApJ...263..835S}. This method has been used in the study of B-type binaries in the 30~Doradus region \citep{2021MNRAS.507.5348V}, resulting in orbital period measurements with relative errors ranging from approximately 0.002\% for VFTS~730 to 0.04\% for VFTS~189. 
In addition, the Heck-Manfroid-Mersch periodogram \citep{1985Heck, 2001Gosset} corrects for the mathematical inaccuracies linked to the Lomb-Scargle method. However overall, all these techniques produce comparable results. 
\newline \indent
Starting from the time, flux, and error on the flux, we calculated the Fourier transform of each subset to determine the dominant frequencies from the light curve. 
These dominant frequencies can be determined with high accuracy due to the long observational baseline with respect to the orbital period that we are trying to detect.
The long time span of our observations ensures that multiple cycles of the periodic signal are observed. Therefore, the peaks corresponding to the true orbital periods are more prominent and distinct in the Fourier transform. 
As it was the case in the O+O systems studied by \citet{Michael2022A&A...666A..18A}, we noticed that, in most cases, the dominant frequency was twice the orbital frequency. This is caused by the symmetric nature of overcontact binary light curves, however, since the orbital periods of our targets have been measured from both photometric and spectroscopic data, we could use the literature values to prevent any confusion. 
\newline \indent
To calculate the error on the orbital frequency and hence on the orbital period, we used the uncertainties incorporated in \textsc{Period04}. More specifically, we used Monte Carlo simulations with 1000 iterations. 
In each iteration, the orbital period was determined from a data set where the times of the data points were the same as for the original input, but the flux was predicted by the best fit plus Gaussian noise scaled to the standard deviation of the data \citep{2005CoAst.146...53L}. These Monte Carlo simulations were then used to determine the uncertainties on the parameter. 
If the error was too large (i.e., larger than $0.2 \%$ for TESS data and larger than $0.002 \%$ for the other databases, based on errors from the literature \citep[e.g.][]{2021MNRAS.507.5348V}) the measurement of the period change rate would be strongly biased and difficult to constrain, so we improved the data set by using more data points and enlarging the subsets that we decided before. Obviously, the period that was determined using \textsc{Period04} is associated with a larger time span of observations. To be able to plot the period in function of the observed time, we decided to assign the calculated orbital period to the mid-time of each observation time span. Once we had all these results, we could start looking at how the period varies in time.

\subsection{Period change determination}
To remain consistent with \citet{Michael2022A&A...666A..18A}, we used a linear fit to characterize the time dependent period evolution. Therefore, the two fitted parameters were the slope and the y intercept, determined with a nonlinear least squares procedure using \texttt{curve\_fit} from SciPy \citep{2020SciPy-NMeth}. To handle all our targets consistently, we decided to offset the time so that the mid-time of each object corresponded to zero on the horizontal axis. 
Additionally, we re-centered the orbital period data before performing the fit by subtracting the known orbital period from literature.
These steps reduce the multicollinearity of the problem by decreasing the correlation between the fitted parameters. The slope is equal to the change in orbital period, $\dot{P}$ and the error is determined based on the covariance matrix. This re-centering does not affect the uncertainty on our final results.
\newline \indent

\section{Results}
\label{sec:results}
\begin{table*}[h]
\caption{Obtained orbital periods for each data set from CT~Tau.}
\label{tab:period}
    \centering
    \begin{tabular}{c P{25mm} P{25mm} c c}
        \hline \hline
         Catalog & First data point \newline [BJD - 2400000] & Central BJD \newline [BJD - 2400000] & $P$ [s] & $P$ [d] \\
         \hline
            DASCH  & 11336  & 14813   & $ 57614.05    \pm 0.04 $   & $0.6668293 \pm 0.0000005 $   \\
            DASCH  & 18292  & 19785   & $ 57614.19    \pm 0.04 $   & $0.6668309 \pm 0.0000005 $   \\
            DASCH  & 21281  & 23053   & $ 57614.10    \pm 0.05 $   & $0.6668287 \pm 0.0000005 $   \\
            DASCH  & 24826  & 25593   & $ 57614.06    \pm 0.11 $   & $0.6668294 \pm 0.0000012 $   \\
            DASCH  & 26361  & 26933   & $ 57614.32    \pm 0.13 $   & $0.6668324 \pm 0.0000015 $   \\
            DASCH  & 27506  & 28394   & $ 57613.82    \pm 0.11 $   & $0.6668266 \pm 0.0000012 $   \\
            DASCH  & 29287  & 29966   & $ 57613.91    \pm 0.14 $   & $0.6668276 \pm 0.0000016 $   \\
            DASCH  & 30652  & 31400   & $ 57614.07    \pm 0.09 $   & $0.6668295 \pm 0.0000010 $   \\
            DASCH  & 32149  & 33489   & $ 57614.01    \pm 0.07 $   & $0.6668289 \pm 0.0000009 $   \\
            DASCH  & 39502  & 43682   & $ 57614.11    \pm 0.04 $    & $0.6668299 \pm 0.0000005 $    \\
            ASAS   & 52622  & 53770   & $ 57613.93    \pm 0.05 $   & $0.6668279 \pm 0.0000006 $   \\
            OMC    & 52692  & 55001   & $ 57613.803   \pm 0.010$   & $0.66682642 \pm 0.00000008$   \\
            OMC    & 57311  & 58571   & $ 57613.796   \pm 0.012$   & $0.66682634 \pm 0.00000014$   \\
            TESS (yr 4)  & 59475  & 59513   & $ 57615.00     \pm 0.11  $ & $0.6668402 \pm 0.0000013  $   \\
         \hline 
    \end{tabular}
    \tablefoot{For each data set, the Barycentric Julian Date (BJD) from the first data point and the central BJD are given.}
\end{table*}
Using all the photometric data mentioned in Sect. \ref{sec:sample}, we are able to determine the orbital periods of each data subset using \textsc{Period04}. 
The long time span of DASCH observations significantly increases the number of available data points, improving the accuracy of our orbital period determinations. 
In previous studies about orbital period change, the stability of the system is expressed using $P/ |\dot{P}|$ which makes it easier to compare different systems with each other. We use CT~Tau as a case study, the analysis and interpretation of the other systems are similar. The results for CT~Tau are listed in Table \ref{tab:period}. The results for all the other binary systems and their catalogs are given in Table \ref{tab:appendix}, while the literature values were given in Table \ref{table:literaturevalues} for comparison. 
\newline \indent
By combining all the period measurements from the archival photometric data, we are able to constrain how quickly the period is changing. All the determined periods are used to find the best linear fit, which in result gives us the slope, a direct measure of $\dot{P}$. Once we obtain $\dot{P}$ and its corresponding error, we calculate the value of $P/|\dot{P}|$. To determine the error on $P/|\dot{P}|$ we use the upper and lower bound values of $\dot{P}$, with this we can calculate the desired asymmetrical errors. In this calculation we neglect the error on the period $P$ as it is orders of magnitude smaller and thus negligible. The results are shown in Table \ref{tab:results} where we show $\dot{P}$ and $P/|\dot{P}|$. Lastly, we also include $\dot{P}/P$ in Table \ref{tab:results}. To visualize our results, the different measurements of the orbital periods obtained from each photometric subset are shown for CT~Tau in Fig. \ref{fig:cttau} and similar figures for the other systems are available in the Appendix \ref{appendixfig}. 

\begin{figure}[ht]
    \centering
    \includegraphics[width=\textwidth/2]{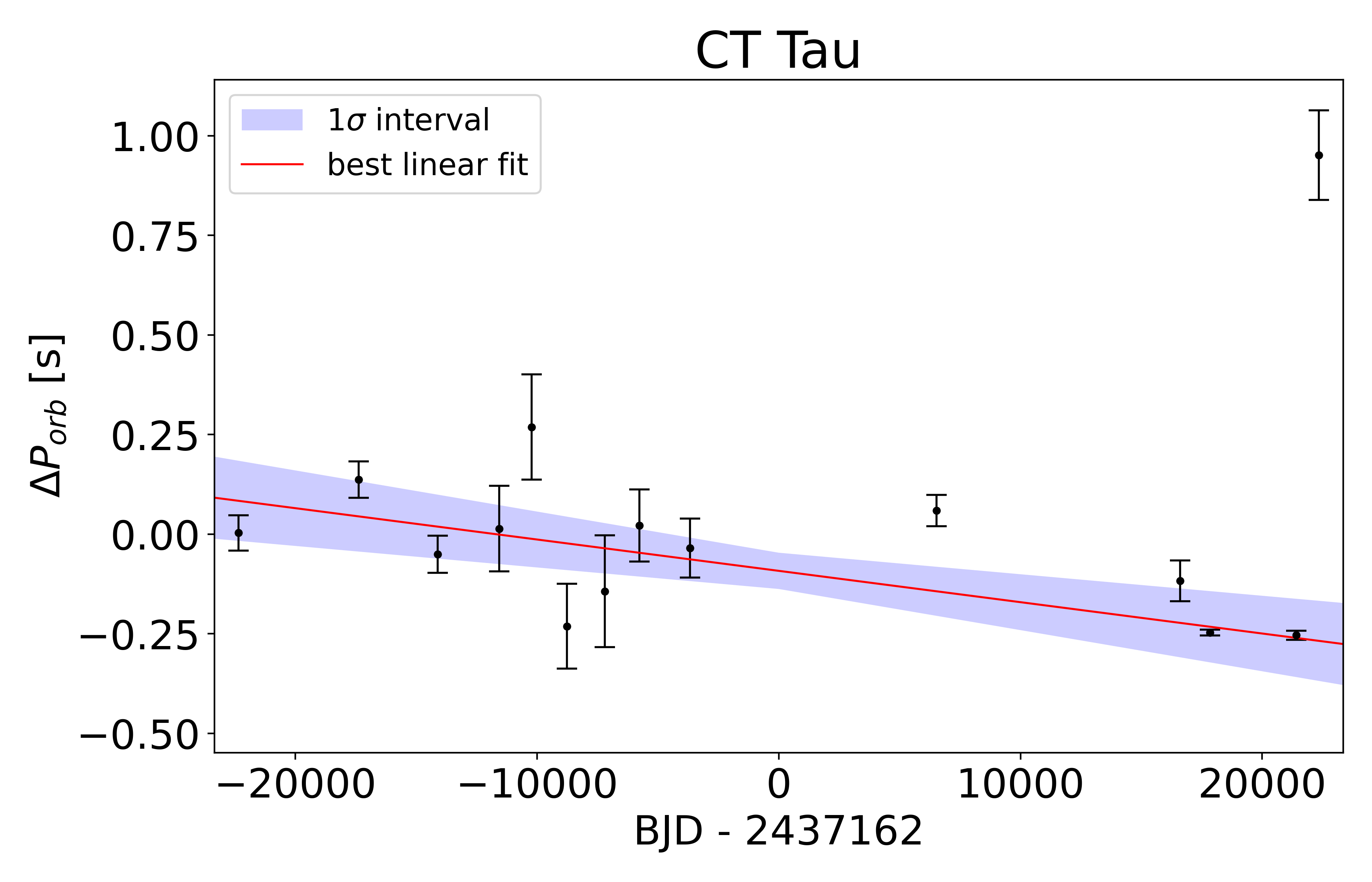}
    \caption{Orbital period change of CT~Tau. On vertical axis, $\Delta P_{orb}$ is the difference between the known orbital period in the literature (see Table \ref{table:literaturevalues}) and the orbital periods determined from archival photometric data of CT~Tau using \textsc{Period04}. The best linear fit is shown in red, and the error region on this fit is pictured in blue.}
    \label{fig:cttau}
\end{figure}
\begin{table*}[ht]
\caption{Results from the linear fit for all our systems.} 
\label{tab:results}
    \centering
    \begin{tabular}{c c c c}
        \hline \hline
         Identifier & $\dot{P}$ [s/yr]& $P/|\dot{P}|$ [Myr]& $\dot{P}/P$ [yr$^{-1}$] \\
         \hline 
         CT~Tau   & $- 0.0029 \pm 0.0009$   & $20 ^{+9}_{-5}$        & $(-5.0 \pm 1.6) \cdot 10^{-8}$       \\
         GU~Mon   & $-0.01 \pm 0.03$        & $10 ^{+\infty}_{-8}$   & $(-1 \pm 4) \cdot 10^{-7}$          \\
         SV~Cen   & $-2.76 \pm 0.06$        & $0.0520 \pm 0.0011 $   & $(-1.92 \pm 0.04) \cdot 10^{-5}$     \\
         V606~Cen & $-0.016 \pm 0.006$      & $8 ^{+5}_{-2}$         & $(-1.3 \pm 0.5) \cdot 10^{-7}$      \\
         V701~Sco & $-0.0095 \pm 0.0006$    & $6.9 ^{+0.5}_{-0.4}$   & $(-1.444 \pm 0.010) \cdot 10^{-7}$ \\
         V745~Cas & $0.13 \pm 0.12$         & $0.9 ^{+8.3}_{-0.4}$   & $(1.1 \pm 1.0) \cdot 10^{-6}$           \\
         VFTS~066 & $-0.8 \pm 0.5$          & $0.07 ^{+0.21}_{-0.05}$&  $(-8 \pm 5) \cdot 10^{-6}$          \\
         \hline
    \end{tabular}
    \tablefoot{The values for $P/|\dot{P}|$ and $\dot{P}/P$ are calculated using $P$ from Table \ref{table:literaturevalues}.}
\end{table*}

From Table \ref{tab:results} we see that $\dot{P}$ is negative for most systems, that almost all have a  $P/ |\dot{P}|$ value between 0.9 and 20 Myr, and that $\dot{P}/P$ are in the range between $10^{-8}$ - $10^{-6}$ [yr$^{-1}$] in most cases. The clear exceptions are SV~Cen and VFTS~066 with the highest $\dot{P}$ and $\dot{P}/P$ values and a $P/ |\dot{P}|$ of around $\sim$ 0.05–0.07 Myr only. 
To understand the timescales, we calculated the nuclear and thermal timescales for the systems in our sample. We found that the nuclear timescale is on the order of a few million years (Myr), while the thermal timescale is on the order of thousands of years (kyr). 
\newline \indent
From Fig. \ref{fig:cttau} and the figures in Appendix \ref{appendixfig}, we see that most of the systems tend to have a slope either negative or close to zero, which means a decreasing or constant orbital period. As mentioned in Sect. \ref{sec:methods}, $\dot{P}$ is determined using a linear fit through our data points. The values in Table \ref{tab:results} are consistent with previous works \citep{Yang2019AJ....157..111Y, Michael2022A&A...666A..18A}. As shown in Table \ref{tab:results}, most of our binary systems except SV~Cen and VFTS~066 have a period stability $P/ |\dot{P}|$  of at least $\sim$1 Myr up to 20 Myr. These time ranges are associated with nuclear timescales, indicating that these systems are evolving on a nuclear timescale. 
%
\newline \indent
For SV~Cen and VFTS~066 the decrease in orbital period is noteworthy and suggests that they are both evolving on a thermal timescale, unlike the five overcontact systems that are evolving on a nuclear timescale. 
Even though SV~Cen is listed in several papers as an overcontact binary \citep{1976MNRAS.176..625W, 1992IAUS..151..369L}, our analysis clearly suggests that it is evolving on a thermal timescale and thereby might indicate that it has a semi-detached configuration, like already suggested by \citet{1994A&A...291..786D, deschamps, davis}. 
For this reason, we will exclude SV~Cen from our discussion about B+B and O+B overcontact binaries. 

\begin{figure*}
    \centering
    \includegraphics[width=0.9\textwidth]{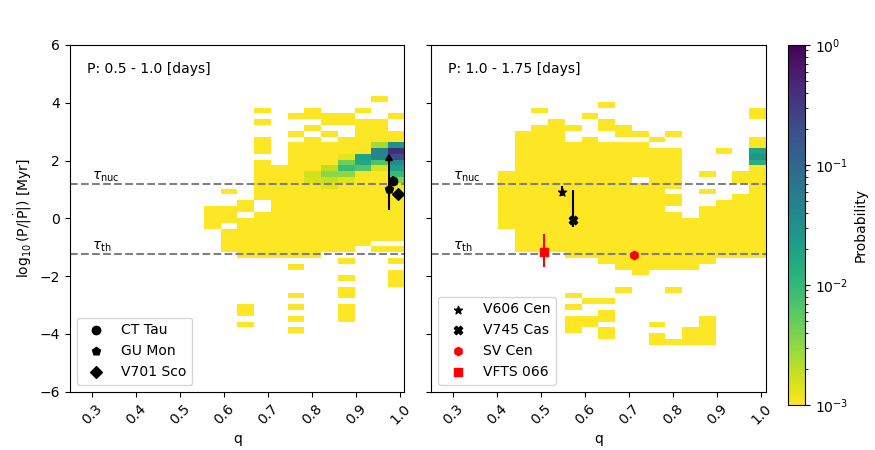}
    \caption{
    Normalized probability distribution based on the population synthesis models from \citet{2021MNRAS.507.5013M}. On the left binned models for an orbital period between 0.5 and 1.0 day, on the right for an orbital period between 1.0 and 1.75 days. On the horizontal axis is the mass ratio $q$ and on the vertical axis $\log_{10}{(P/|\dot{P}|)}$. The colors indicate, in log scale, the probability of having a system with those values of $q$ and $\log_{10}{(P/|\dot{P}|)}$. Anything with a probability of $\leq10^{-3}$ are indicated in yellow. In black the studied systems are located on the probability distribution plot, with their corresponding errors. If the value does not have an upper limit, we indicate this error with an arrow. The red-colored binary systems are those excluded from the discussion about overcontact binaries. For reference, we plot the nuclear and thermal time scales of a 12M$_{\odot}$ star calculated based on evolutionary tracks from \citet{Brott2011} in gray. }
    \label{population}
\end{figure*}
\begin{figure*}
    \centering
    \includegraphics[width=0.9\textwidth]{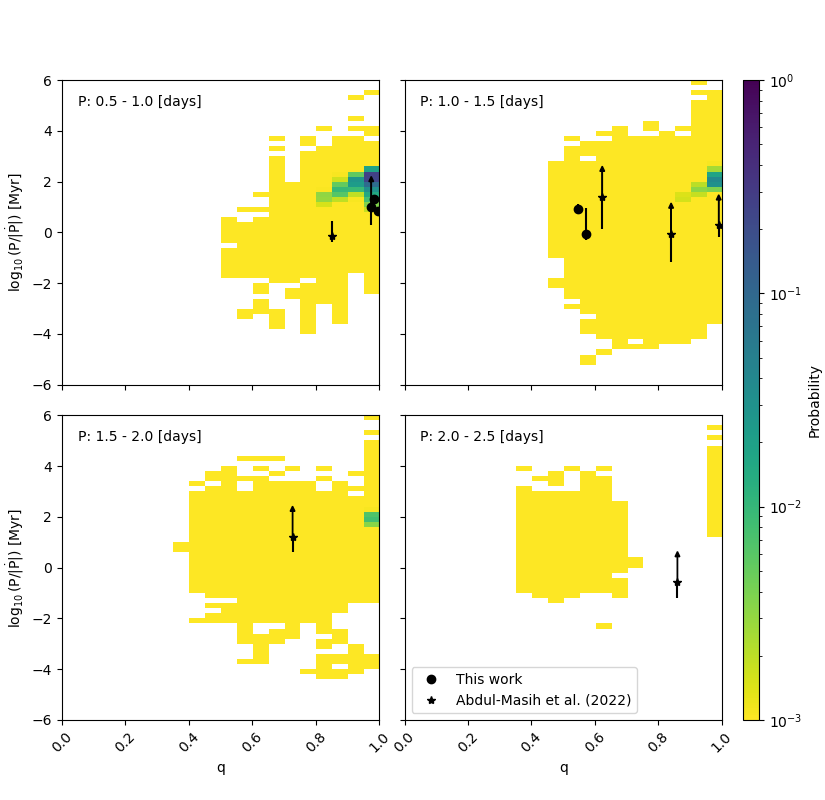}
    \caption{Normalized probability distribution in a similar figure like Fig. \ref{population} binned for an orbital period between 0.5 days and 2.5 days. For comparison, our final five studied B+B overcontact systems and the previously studied O+O overcontact systems from \citet{Michael2022A&A...666A..18A} are shown.}
    \label{population_all}
\end{figure*}
Concerning VFTS~066, it is important to mention that the fit was constituted of only three data points with relatively large errors. In the photometric analysis of \citet{2020AA...634A.119M} an inclination of $17.5 ^{+3.2}_{-2.5}$ $^\circ$ was found. At such a low inclination, the system is not eclipsing and therefore the light curve is dominated by ellipsoidal variations, which would appear quite similar for overcontact and semi-detached configurations. Additionally, the best fit solution (while favoring an overcontact configuration) fell quite close to the border between the two regimes. Given this and the fact that the period stability is significantly lower than the rest of the sample, it is possible that the system was identified as an overcontact system while actually being (semi-)detached or close to contact. In addition, with the timescale derived from $\dot{P}$ we will exclude VFTS~066 from the discussion about overcontact binaries. 
 

\section{Discussion}
\label{sec:discuss}
We compare our observations with the population synthesis simulations of \citet{2021MNRAS.507.5013M}, similar to the analysis done in \citet{Michael2022A&A...666A..18A} for O+O overcontact binaries. 
The binary models in the population synthesis are initialized with total masses ranging from 20$M_\odot$ to 80$M_\odot$, initial mass ratios from 0.6 to 1 and initial orbital periods from 0.6 days to 1 day. These initial values represent the parameters of the binary system at the beginning of the zero age main sequence (ZAMS). 
These models focus only on the main sequence phase of the binary systems. 
For the population distribution, the study examined the fraction of time these systems spend in specific configurations, defined by combinations of their current total mass, mass ratio and orbital period.
Using these results, the population distribution predicts the configurations in which these binary systems are most likely to spend the majority of their lifetimes and be detected in observations.
\newline \indent
The probability plots based on the models of \citet{2021MNRAS.507.5013M} are shown in Fig. \ref{population} with the colors in log scale. In the left-hand panel, we use the models with a current orbital period between 0.5 and 1.0 days, while the right panel shows periods between 1.0 and 1.75 days. 
It shows the normalized theoretical probability distribution of $P/|\dot{P}|$ as a function of mass ratio. Similar to the O+O models as shown in Fig. \ref{population_all}, we notice that for B+B systems with orbital periods of $P \leq 1.75$ days, the probability peaks at a mass ratio $q=1$. 
However, for longer orbital periods, the distribution appears to flatten, and $q < 1$ becomes more likely. 
\newline \indent
From the studied sample of systems with an orbital period between 0.5 and 1.0 day, the sample has, indeed, a mass ratio close to 1, however, the values of $\log_{10}{(P/|\dot{P}|)}$ are slightly lower than the peak of the models' distribution. 
Nonetheless, these B+B overcontact systems are in better agreement with the models of \citet{2021MNRAS.507.5013M} compared to the O+O overcontact sample. 
To illustrate this even better, we additionally showed both samples in Fig. \ref{population_all} for comparison. 
This trend of mass ratios close to 1 is not visible for the systems with a longer orbital period. In comparison with the models, V606~Cen and V745~Cas are found more in the tail of the probability distribution, with a lower mass ratio $q$. 
For systems with an orbital period $P \geq 1.75$ days, we find the same discrepancy between our observations and the predictions as concluded in \citet{2021MNRAS.507.5013M} and \citet{Michael2022A&A...666A..18A}, which are based on the models from the population synthesis. See Fig. \ref{population_all} for reference. 
\newline \indent
We still need to exercise caution regarding this discrepancy, as multiple factors could have influenced our results. Possible factors include differences in metallicity assumptions. Notably, the models of \citet{2021MNRAS.507.5013M} were conducted at Large Magellanic Cloud metallicity, potentially leading to variations compared to Galactic systems. Apart from metallicity considerations, limitations in the orbital period and mass transfer assumptions, constraints on initial mass ratios, absence of energy transfer considerations \citep{fabry, fabry2}, and limitations in the contact scheme in MESA may also contribute to discrepancies. A more elaborate discussion about these aspects can be found in \citet{Michael2022A&A...666A..18A}. From the population synthesis plots we see that we find four in the yellow and one in the blue regions, this means that the observations are not distributed in the same way as the population synthesis predicts. 
\newline \indent
The objects in the blue region should outnumber those in the yellow region by a ratio of at least 1000 to 1. While the yellow regions, representing probabilities smaller than $10^{-3}$, are visible on the graph, differences within the yellow regions are not readily apparent. However, a trend toward higher mass ratios is still present.
While there is an acceptable correspondence in the 0.5–1.0 orbital period bin, the identification of two poorly matching objects in the 1.0–1.75 orbital period bin signifies dissimilarity in the distributions.
\newline \indent
Our sample containing five B+B overcontact binaries has given us more insight in the possible discrepancy between theoretical models and the observations, considering the timescale of equalization of binaries. To strengthen these results and formulate a definite conclusion, a larger sample of massive overcontact binaries is needed to better investigate their orbital period changes. 

\section{Conclusion}
\label{sec:concl}
Given the discrepancy between the theoretical predictions and observations presented in \citet{Michael2022A&A...666A..18A}, we set out to both provide additional observational constraints and verify this discrepancy by analyzing the period stability of a sample of B+B and O+B overcontact systems. We know B-type stars are more numerous than O-type stars, yet, the amount of known B+B overcontact binary systems is still limited. We included every system that we found in literature that is characterized as a B+B and O+B binary in the overcontact phase and fit our selection criteria. This gave us a sample of five overcontact systems and two impostors that were investigated using their available photometric data. 
\newline \indent
Using archival photometric data spanning back as far as a century, we measured the orbital period using \textsc{Period04} of each system using subsets of the data in order to study the change of this parameter with time. We notice that the choice of data subset had a great influence on the end results of \textsc{Period04}. In most cases, more data points will give a smaller error on the orbital period and make the linear fit to determine the period change much more challenging. Therefore, the selection and division of the data sets are not absolute but rather subject to the researcher's choices. To minimize these influences, the larger data sets are split up in data subsets of approximately 10 years. Only, when there are obvious gaps or differences, we have split up the data sets manually. 
\newline \indent
Subsequently, using a linear fit we were able to deduce a quantitative value for the period variations, that is $\dot{P}$, hence $P/|\dot{P}|$ and $\dot{P}/P$. Based on these results, a significant divergence was observed in the data from SV~Cen and VFTS~066 compared to our other targets, suggesting its evolution on a thermal timescale. Even though the overcontact binary literature often lists SV~Cen as such, we also found papers suggesting that SV~Cen is an Algol system \citep{2011A&A...528A..16V}, but after measuring its period change, we tend to support the latter and decided to exclude SV~Cen from our sample of massive overcontact binaries. VFTS~066's limited data points, combined with its small inclination and therefore dominating ellipsoidal variations in the light curve, suggest caution in its classification as an overcontact system. Therefore, it is excluded as well from our sample.
Among the remaining five systems, we find that $P/|\dot{P}|$ falls between $\sim$1 and 20 Myr, indicating an evolution on a nuclear timescale. Comparing our results with population synthesis simulations, we observed similar discrepancies, particularly for $1.0 \leq P \leq 1.75$ days, between the predicted and observed distributions, as seen in a previous study of O+O massive overcontact systems. 
\newline \indent
In conclusion, this study has provided new insights towards the understanding of orbital period variations and evolutionary timescales of a diverse set of B+B massive overcontact binary systems. The observed discrepancies in the mass ratio distribution for B+B overcontact binaries between empirical data and theoretical predictions from population synthesis models highlight the need for further refinement. Enhancing these models is crucial to capture the diverse nature of overcontact binaries in various evolutionary stages more accurately. This also highlights the importance of expanding the sample size for a more comprehensive understanding. This research, therefore, serves as a valuable cornerstone for future studies aimed at clarifying the evolution of overcontact binary systems.

\begin{acknowledgements}
We acknowledge support from the ESO Science Support Discretionary Fund under program ID SSDF 17/23 C.
This project received the support of two fellowships from ”La Caixa” Foundation (ID 100010434). The fellowship codes are LCF/BQ/PI23/11970035 (MAM) and LCF/BQ/PI23/11970031 (AE).
PM acknowledges support from the FWO senior postdoctoral fellowship No. 12ZY523N
This research has made use of the ExoDat Database, operated at LAM-OAMP, Marseille, France, on behalf of the CoRoT/Exoplanet program.
Based on data from the OMC Archive at CAB (INTA-CSIC), pre-processed by ISDC and further processed by the OMC Team at CAB.
The OMC Archive is part of the Spanish Virtual Observatory project. Both are funded by MCIN/AEI/10.13039/501100011033 through grants PID2020-112949GB-I00 and PID2019-107061GB-C61, respectively.
This paper includes data collected by the TESS mission, which are publicly available from the Mikulski Archive for Space Telescopes (MAST). Funding for the TESS mission is provided by NASA’s Science Mission directorate.
This research made use of Lightkurve, a Python package for Kepler and TESS data analysis (Lightkurve Collaboration, 2018).
The DASCH project at Harvard is grateful for partial support from NSF grants AST-0407380, AST-0909073, and AST-1313370

\end{acknowledgements}

\bibliographystyle{aa} 
\bibliography{bibtex.bib} 

\begin{thebibliography}{84}
\expandafter\ifx\csname natexlab\endcsname\relax\def\natexlab#1{#1}\fi

\bibitem[{{Abdul-Masih} {et~al.}(2022){Abdul-Masih}, {Escorza}, {Menon},
  {Mahy}, \& {Marchant}}]{Michael2022A&A...666A..18A}
{Abdul-Masih}, M., {Escorza}, A., {Menon}, A., {Mahy}, L., \& {Marchant}, P.
  2022, \aap, 666, A18

\bibitem[{{Abdul-Masih} {et~al.}(2020){Abdul-Masih}, {Sana}, {Conroy},
  {Sundqvist}, {Pr{\v{s}}a}, {Kochoska}, \& {Puls}}]{2020A&A...636A..59A}
{Abdul-Masih}, M., {Sana}, H., {Conroy}, K.~E., {et~al.} 2020, \aap, 636, A59

\bibitem[{{Abdul-Masih} {et~al.}(2021){Abdul-Masih}, {Sana}, {Hawcroft},
  {Almeida}, {Brands}, {de Mink}, {Justham}, {Langer}, {Mahy}, {Marchant},
  {Menon}, {Puls}, \& {Sundqvist}}]{Michael2021A&A...651A..96A}
{Abdul-Masih}, M., {Sana}, H., {Hawcroft}, C., {et~al.} 2021, \aap, 651, A96

\bibitem[{Alfonso-Garzón {et~al.}(2012)Alfonso-Garzón, Domingo, Mas-Hesse, \&
  Giménez}]{omc}
Alfonso-Garzón, J., Domingo, A., Mas-Hesse, J.~M., \& Giménez, A. 2012,
  Astronomy and Astrophysics, 548, A79

\bibitem[{{Andersen} {et~al.}(1980){Andersen}, {Nordstrom}, \&
  {Wilson}}]{1980A&A....82..225A}
{Andersen}, J., {Nordstrom}, B., \& {Wilson}, R.~E. 1980, \aap, 82, 225

\bibitem[{{Applegate} \& {Shaham}(1994)}]{1994ApJ...436..312A}
{Applegate}, J.~H. \& {Shaham}, J. 1994, \apj, 436, 312

\bibitem[{{Barentsen} \& {Lightkurve
  Collaboration}(2020)}]{2020AAS...23540904B}
{Barentsen}, G. \& {Lightkurve Collaboration}. 2020, in American Astronomical
  Society Meeting Abstracts, Vol. 235, American Astronomical Society Meeting
  Abstracts \#235, 409.04

\bibitem[{{Bell} \& {Malcolm}(1987)}]{1987MNRAS.226..899B}
{Bell}, S.~A. \& {Malcolm}, G.~J. 1987, \mnras, 226, 899

\bibitem[{{Bobylev} \& {Bajkova}(2015)}]{2015AstL...41..473B}
{Bobylev}, V.~V. \& {Bajkova}, A.~T. 2015, Astronomy Letters, 41, 473

\bibitem[{{Brott} {et~al.}(2011){Brott}, {de Mink}, {Cantiello}, {Langer}, {de
  Koter}, {Evans}, {Hunter}, {Trundle}, \& {Vink}}]{Brott2011}
{Brott}, I., {de Mink}, S.~E., {Cantiello}, M., {et~al.} 2011, \aap, 530, A115

\bibitem[{{Davis} {et~al.}(2014){Davis}, {Siess}, \& {Deschamps}}]{davis}
{Davis}, P.~J., {Siess}, L., \& {Deschamps}, R. 2014, \aap, 570, A25

\bibitem[{{De Greve} \& {Linnell}(1994)}]{1994A&A...291..786D}
{De Greve}, J.~P. \& {Linnell}, A.~P. 1994, \aap, 291, 786

\bibitem[{{Deleuil} {et~al.}(2009){Deleuil}, {Meunier}, {Moutou}, {Surace},
  {Deeg}, {Barbieri}, {Debosscher}, {Almenara}, {Agneray}, {Granet},
  {Guterman}, \& {Hodgkin}}]{CoRoT2009AJ....138..649D}
{Deleuil}, M., {Meunier}, J.~C., {Moutou}, C., {et~al.} 2009, \aj, 138, 649

\bibitem[{{Deschamps} {et~al.}(2013){Deschamps}, {Siess}, {Davis}, \&
  {Jorissen}}]{deschamps}
{Deschamps}, R., {Siess}, L., {Davis}, P.~J., \& {Jorissen}, A. 2013, \aap,
  557, A40

\bibitem[{{Drechsel}(1994{\natexlab{a}})}]{1994AGAb...10...95D}
{Drechsel}, H. 1994{\natexlab{a}}, in Astronomische Gesellschaft Abstract
  Series, Vol.~10, Astronomische Gesellschaft Abstract Series, 95--95

\bibitem[{{Drechsel} \& {Lorenz}(1993)}]{1993IBVS.3868....1D}
{Drechsel}, H. \& {Lorenz}, R. 1993, Information Bulletin on Variable Stars,
  3868, 1

\bibitem[{{Drechsel} {et~al.}(1982){Drechsel}, {Rahe}, {Wargau}, \&
  {Wolf}}]{1982A&A...110..246D}
{Drechsel}, H., {Rahe}, J., {Wargau}, W., \& {Wolf}, B. 1982, \aap, 110, 246

\bibitem[{{Drechsel}(1994{\natexlab{b}})}]{1994iue..prop.4773D}
{Drechsel}, H.~J. 1994{\natexlab{b}}, {The Evolutionary State of SV Centauri},
  IUE Proposal ID SVPHD

\bibitem[{{Eggen}(1961)}]{1961RGOB...27...61E}
{Eggen}, O.~J. 1961, Royal Greenwich Observatory Bulletins, 27, 61

\bibitem[{{Fabry} {et~al.}(2023){Fabry}, {Marchant}, {Langer}, \&
  {Sana}}]{fabry2}
{Fabry}, M., {Marchant}, P., {Langer}, N., \& {Sana}, H. 2023, \aap, 672, A175

\bibitem[{{Fabry} {et~al.}(2022){Fabry}, {Marchant}, \& {Sana}}]{fabry}
{Fabry}, M., {Marchant}, P., \& {Sana}, H. 2022, \aap, 661, A123

\bibitem[{{Franco}(1994)}]{1994A&AS..104....9F}
{Franco}, G.~A.~P. 1994, \aaps, 104, 9

\bibitem[{Gosset {et~al.}(2001)Gosset, Royer, Rauw, Manfroid, \&
  Vreux}]{2001Gosset}
Gosset, E., Royer, P., Rauw, G., Manfroid, J., \& Vreux, J.-M. 2001, Monthly
  Notices of the Royal Astronomical Society, 327, 435

\bibitem[{{Grindlay}(2017)}]{dasch}
{Grindlay}, J. 2017, in Astrophysics and Space Science Proceedings, Vol.~50,
  The Science of Time 2016, ed. E.~F. {Arias}, L.~{Combrinck}, P.~{Gabor},
  C.~{Hohenkerk}, \& P.~K. {Seidelmann}, 203

\bibitem[{{Grudi{\'c}} \& {Hopkins}(2019)}]{grudic}
{Grudi{\'c}}, M.~Y. \& {Hopkins}, P.~F. 2019, \mnras, 488, 2970

\bibitem[{{Heck} {et~al.}(1985){Heck}, {Manfroid}, \& {Mersch}}]{1985Heck}
{Heck}, A., {Manfroid}, J., \& {Mersch}, G. 1985, \aaps, 59, 63

\bibitem[{{Heger} {et~al.}(2003){Heger}, {Fryer}, {Woosley}, {Langer}, \&
  {Hartmann}}]{heger}
{Heger}, A., {Fryer}, C.~L., {Woosley}, S.~E., {Langer}, N., \& {Hartmann},
  D.~H. 2003, \apj, 591, 288

\bibitem[{{Heger} {et~al.}(2023){Heger}, {M{\"u}ller}, \& {Mandel}}]{heger2023}
{Heger}, A., {M{\"u}ller}, B., \& {Mandel}, I. 2023, arXiv e-prints,
  arXiv:2304.09350

\bibitem[{{Henneco} {et~al.}(2024){Henneco}, {Schneider}, \&
  {Laplace}}]{2024A&A...682A.169H}
{Henneco}, J., {Schneider}, F.~R.~N., \& {Laplace}, E. 2024, \aap, 682, A169

\bibitem[{{Herczeg} \& {Drechsel}(1985)}]{1985Ap&SS.114....1H}
{Herczeg}, T.~J. \& {Drechsel}, H. 1985, \apss, 114, 1

\bibitem[{{Herrero}(2008)}]{herrero}
{Herrero}, A. 2008, in Revista Mexicana de Astronomia y Astrofisica Conference
  Series, Vol.~33, Revista Mexicana de Astronomia y Astrofisica Conference
  Series, 15--22

\bibitem[{{Hurley} {et~al.}(2002){Hurley}, {Tout}, \& {Pols}}]{hurley}
{Hurley}, J.~R., {Tout}, C.~A., \& {Pols}, O.~R. 2002, \mnras, 329, 897

\bibitem[{{Hut}(1981)}]{hut}
{Hut}, P. 1981, \aap, 99, 126

\bibitem[{{Irwin} \& {Landolt}(1972)}]{1972PASP...84..686I}
{Irwin}, J.~B. \& {Landolt}, A.~U. 1972, \pasp, 84, 686

\bibitem[{{Istchenko} \& {Chugainov}(1965)}]{1965IBVS...95....1I}
{Istchenko}, I.~M. \& {Chugainov}, P.~F. 1965, Information Bulletin on Variable
  Stars, 95, 1

\bibitem[{{Jenkins} {et~al.}(2021){Jenkins}, {Twicken}, {Caldwell}, {Ting},
  {Tenenbaum}, {Smith}, {Wohler}, {Rose}, {Henze}, {Fausnaugh}, {Burke},
  {Vanderspek}, \& {Hounsell}}]{spoc}
{Jenkins}, J.~M., {Twicken}, J.~D., {Caldwell}, D.~A., {et~al.} 2021, in
  Posters from the TESS Science Conference II (TSC2), 183

\bibitem[{{Langer}(2012)}]{2012ARA&A..50..107L}
{Langer}, N. 2012, \araa, 50, 107

\bibitem[{{Lee} {et~al.}(1991){Lee}, {Osaki}, \& {Saio}}]{Lee1991}
{Lee}, U., {Osaki}, Y., \& {Saio}, H. 1991, \mnras, 250, 432

\bibitem[{{Lenz} \& {Breger}(2005)}]{2005CoAst.146...53L}
{Lenz}, P. \& {Breger}, M. 2005, Communications in Asteroseismology, 146, 53

\bibitem[{{Leung}(1974)}]{1974A&AS...13..315L}
{Leung}, K.~C. 1974, \aaps, 13, 315

\bibitem[{{Li} {et~al.}(2022){Li}, {Liao}, {Qian}, {Fern{\'a}ndez Laj{\'u}s},
  {Zhang}, \& {Zhao}}]{Li2022ApJ...924...30L}
{Li}, F.~X., {Liao}, W.~P., {Qian}, S.~B., {et~al.} 2022, \apj, 924, 30

\bibitem[{{Linnell}(1992)}]{1992IAUS..151..369L}
{Linnell}, A.~P. 1992, in Evolutionary Processes in Interacting Binary Stars,
  ed. Y.~{Kondo}, R.~{Sistero}, \& R.~S. {Polidan}, Vol. 151, 369

\bibitem[{{Lomb}(1976)}]{1976Ap&SS..39..447L}
{Lomb}, N.~R. 1976, \apss, 39, 447

\bibitem[{{Lorenz} {et~al.}(1999){Lorenz}, {Mayer}, \&
  {Drechsel}}]{1999A&A...345..531L}
{Lorenz}, R., {Mayer}, P., \& {Drechsel}, H. 1999, \aap, 345, 531

\bibitem[{{Lorenzo} {et~al.}(2016){Lorenzo}, {Negueruela}, {Vilardell},
  {Sim{\'o}n-D{\'\i}az}, {Pastor}, \& {M{\'e}ndez Majuelos}}]{Lorenzo2016}
{Lorenzo}, J., {Negueruela}, I., {Vilardell}, F., {et~al.} 2016, \aap, 590, A45

\bibitem[{{Maeda}(2022)}]{maeda}
{Maeda}, K. 2022, in Handbook of X-ray and Gamma-ray Astrophysics, 75

\bibitem[{Maeder \& Meynet(2012)}]{RevModPhys.84.25}
Maeder, A. \& Meynet, G. 2012, Rev. Mod. Phys., 84, 25

\bibitem[{{Mahy} {et~al.}(2020{\natexlab{a}}){Mahy}, {Almeida}, {Sana},
  {Clark}, {de Koter}, {de Mink}, {Evans}, {Grin}, {Langer}, {Moffat},
  {Schneider}, {Shenar}, \& {Tramper}}]{2020AA...634A.119M}
{Mahy}, L., {Almeida}, L.~A., {Sana}, H., {et~al.} 2020{\natexlab{a}}, \aap,
  634, A119

\bibitem[{{Mahy} {et~al.}(2020{\natexlab{b}}){Mahy}, {Sana}, {Abdul-Masih},
  {Almeida}, {Langer}, {Shenar}, {de Koter}, {de Mink}, {de Wit}, {Grin},
  {Evans}, {Moffat}, {Schneider}, {Barb{\'a}}, {Clark}, {Crowther},
  {Gr{\"a}fener}, {Lennon}, {Tramper}, \& {Vink}}]{2020A&A...634A.118M}
{Mahy}, L., {Sana}, H., {Abdul-Masih}, M., {et~al.} 2020{\natexlab{b}}, \aap,
  634, A118

\bibitem[{{Marchant} \& {Bodensteiner}(2023)}]{annualrevpablo}
{Marchant}, P. \& {Bodensteiner}, J. 2023, arXiv e-prints, arXiv:2311.01865

\bibitem[{{McLeod} {et~al.}(2019){McLeod}, {Dale}, {Evans}, {Ginsburg},
  {Kruijssen}, {Pellegrini}, {Ramsay}, \& {Testi}}]{mcleod}
{McLeod}, A.~F., {Dale}, J.~E., {Evans}, C.~J., {et~al.} 2019, \mnras, 486,
  5263

\bibitem[{{Menon} {et~al.}(2021){Menon}, {Langer}, {de Mink}, {Justham}, {Sen},
  {Sz{\'e}csi}, {de Koter}, {Abdul-Masih}, {Sana}, {Mahy}, \&
  {Marchant}}]{2021MNRAS.507.5013M}
{Menon}, A., {Langer}, N., {de Mink}, S.~E., {et~al.} 2021, \mnras, 507, 5013

\bibitem[{{Moe} \& {Di Stefano}(2017)}]{Moe2017}
{Moe}, M. \& {Di Stefano}, R. 2017, \apjs, 230, 15

\bibitem[{{Mohamed} \& {Podsiadlowski}(2007)}]{Mohamed2007}
{Mohamed}, S. \& {Podsiadlowski}, P. 2007, in Astronomical Society of the
  Pacific Conference Series, Vol. 372, 15th European Workshop on White Dwarfs,
  ed. R.~{Napiwotzki} \& M.~R. {Burleigh}, 397

\bibitem[{{Nakamura} \& {Nakamura}(1982)}]{1982Ap&SS..83..163N}
{Nakamura}, M. \& {Nakamura}, Y. 1982, \apss, 83, 163

\bibitem[{{Paczy{\'n}ski}(1971)}]{paczynski}
{Paczy{\'n}ski}, B. 1971, \araa, 9, 183

\bibitem[{{Parsons} {et~al.}(2010){Parsons}, {Marsh}, {Copperwheat}, {Dhillon},
  {Littlefair}, {Hickman}, {Maxted}, {G{\"a}nsicke}, {Unda-Sanzana}, {Colque},
  {Barraza}, {S{\'a}nchez}, \& {Monard}}]{2010MNRAS.407.2362P}
{Parsons}, S.~G., {Marsh}, T.~R., {Copperwheat}, C.~M., {et~al.} 2010, \mnras,
  407, 2362

\bibitem[{{Paxton} {et~al.}(2011){Paxton}, {Bildsten}, {Dotter}, {Herwig},
  {Lesaffre}, \& {Timmes}}]{mesa1}
{Paxton}, B., {Bildsten}, L., {Dotter}, A., {et~al.} 2011, \apjs, 192, 3

\bibitem[{{Paxton} {et~al.}(2013){Paxton}, {Cantiello}, {Arras}, {Bildsten},
  {Brown}, {Dotter}, {Mankovich}, {Montgomery}, {Stello}, {Timmes}, \&
  {Townsend}}]{mesa2}
{Paxton}, B., {Cantiello}, M., {Arras}, P., {et~al.} 2013, \apjs, 208, 4

\bibitem[{{Paxton} {et~al.}(2015){Paxton}, {Marchant}, {Schwab}, {Bauer},
  {Bildsten}, {Cantiello}, {Dessart}, {Farmer}, {Hu}, {Langer}, {Townsend},
  {Townsley}, \& {Timmes}}]{mesa3}
{Paxton}, B., {Marchant}, P., {Schwab}, J., {et~al.} 2015, \apjs, 220, 15

\bibitem[{{Perryman} {et~al.}(1997){Perryman}, {Lindegren}, {Kovalevsky},
  {Hoeg}, {Bastian}, {Bernacca}, {Cr{\'e}z{\'e}}, {Donati}, {Grenon},
  {Grewing}, {van Leeuwen}, {van der Marel}, {Mignard}, {Murray}, {Le Poole},
  {Schrijver}, {Turon}, {Arenou}, {Froeschl{\'e}}, \&
  {Petersen}}]{1997A&A...323L..49P}
{Perryman}, M.~A.~C., {Lindegren}, L., {Kovalevsky}, J., {et~al.} 1997, \aap,
  323, L49

\bibitem[{{Plaut}(1948)}]{1948AnLei..20....3P}
{Plaut}, L. 1948, Annalen van de Sterrewacht te Leiden, 20, 3

\bibitem[{{Plewa} \& {Wlodarczyk}(1993)}]{Plewa1993}
{Plewa}, T. \& {Wlodarczyk}, K.~J. 1993, \actaa, 43, 249

\bibitem[{{Pojmanski}(1997)}]{asas}
{Pojmanski}, G. 1997, \actaa, 47, 467

\bibitem[{{Pojmanski}(2003)}]{2003AcA....53..341P}
{Pojmanski}, G. 2003, \actaa, 53, 341

\bibitem[{{Pols}(1994)}]{1994A&A...290..119P}
{Pols}, O.~R. 1994, \aap, 290, 119

\bibitem[{{Ricker} {et~al.}(2015){Ricker}, {Winn}, {Vanderspek}, {Latham},
  {Bakos}, {Bean}, {Berta-Thompson}, {Brown}, {Buchhave}, {Butler}, {Butler},
  {Chaplin}, {Charbonneau}, {Christensen-Dalsgaard}, {Clampin}, {Deming},
  {Doty}, {De Lee}, {Dressing}, {Dunham}, {Endl}, {Fressin}, {Ge}, {Henning},
  {Holman}, {Howard}, {Ida}, {Jenkins}, {Jernigan}, {Johnson}, {Kaltenegger},
  {Kawai}, {Kjeldsen}, {Laughlin}, {Levine}, {Lin}, {Lissauer}, {MacQueen},
  {Marcy}, {McCullough}, {Morton}, {Narita}, {Paegert}, {Palle}, {Pepe},
  {Pepper}, {Quirrenbach}, {Rinehart}, {Sasselov}, {Sato}, {Seager},
  {Sozzetti}, {Stassun}, {Sullivan}, {Szentgyorgyi}, {Torres}, {Udry}, \&
  {Villasenor}}]{TESS}
{Ricker}, G.~R., {Winn}, J.~N., {Vanderspek}, R., {et~al.} 2015, Journal of
  Astronomical Telescopes, Instruments, and Systems, 1, 014003

\bibitem[{{Rucinski} {et~al.}(1992){Rucinski}, {Baade}, {Lu}, \&
  {Udalski}}]{1992AJ....103..573R}
{Rucinski}, S.~M., {Baade}, D., {Lu}, W.~X., \& {Udalski}, A. 1992, \aj, 103,
  573

\bibitem[{{Sana} {et~al.}(2012){Sana}, {de Mink}, {de Koter}, {Langer},
  {Evans}, {Gieles}, {Gosset}, {Izzard}, {Le Bouquin}, \& {Schneider}}]{sana}
{Sana}, H., {de Mink}, S.~E., {de Koter}, A., {et~al.} 2012, Science, 337, 444

\bibitem[{{Scargle}(1982)}]{1982ApJ...263..835S}
{Scargle}, J.~D. 1982, \apj, 263, 835

\bibitem[{{Shu} {et~al.}(1979){Shu}, {Lubow}, \& {Anderson}}]{shufh}
{Shu}, F.~H., {Lubow}, S.~H., \& {Anderson}, L. 1979, \apj, 229, 223

\bibitem[{Szymanski(2006)}]{szymanski2006optical}
Szymanski, M.~K. 2006, The Optical Gravitational Lensing Experiment. Internet
  Access to the OGLE Photometry Data Set: OGLE-II BVI maps and I-band data

\bibitem[{{van Leeuwen}(2009)}]{2009A&A...500..505V}
{van Leeuwen}, F. 2009, \aap, 500, 505

\bibitem[{{van Rensbergen} {et~al.}(2011){van Rensbergen}, {de Greve},
  {Mennekens}, {Jansen}, \& {de Loore}}]{2011A&A...528A..16V}
{van Rensbergen}, W., {de Greve}, J.~P., {Mennekens}, N., {Jansen}, K., \& {de
  Loore}, C. 2011, \aap, 528, A16

\bibitem[{{Villase{\~n}or} {et~al.}(2021){Villase{\~n}or}, {Taylor}, {Evans},
  {Ram{\'\i}rez-Agudelo}, {Sana}, {Almeida}, {de Mink}, {Dufton}, \&
  {Langer}}]{2021MNRAS.507.5348V}
{Villase{\~n}or}, J.~I., {Taylor}, W.~D., {Evans}, C.~J., {et~al.} 2021,
  \mnras, 507, 5348

\bibitem[{Virtanen {et~al.}(2020)Virtanen, Gommers, Oliphant, Haberland, Reddy,
  Cournapeau, Burovski, Peterson, Weckesser, Bright, {van der Walt}, Brett,
  Wilson, Millman, Mayorov, Nelson, Jones, Kern, Larson, Carey, Polat, Feng,
  Moore, {VanderPlas}, Laxalde, Perktold, Cimrman, Henriksen, Quintero, Harris,
  Archibald, Ribeiro, Pedregosa, {van Mulbregt}, \& {SciPy 1.0
  Contributors}}]{2020SciPy-NMeth}
Virtanen, P., Gommers, R., Oliphant, T.~E., {et~al.} 2020, Nature Methods, 17,
  261

\bibitem[{{Wellstein} {et~al.}(2001){Wellstein}, {Langer}, \&
  {Braun}}]{2001A&A...369..939W}
{Wellstein}, S., {Langer}, N., \& {Braun}, H. 2001, \aap, 369, 939

\bibitem[{{Wilson}(1979)}]{wilson}
{Wilson}, R.~E. 1979, \apj, 234, 1054

\bibitem[{{Wilson} \& {Starr}(1976)}]{1976MNRAS.176..625W}
{Wilson}, R.~E. \& {Starr}, T.~C. 1976, \mnras, 176, 625

\bibitem[{{Woosley} {et~al.}(2002){Woosley}, {Heger}, \& {Weaver}}]{woosley}
{Woosley}, S.~E., {Heger}, A., \& {Weaver}, T.~A. 2002, Reviews of Modern
  Physics, 74, 1015

\bibitem[{Wu {et~al.}(2023)Wu, Qian, Li, Zejda, Mikulásek, Zhu, Liao, \&
  Zhao}]{Wu10.1093/pasj/psad003}
Wu, C., Qian, S., Li, F., {et~al.} 2023, Publications of the Astronomical
  Society of Japan, 75, 358

\bibitem[{{Yang} {et~al.}(2019){Yang}, {Yuan}, \&
  {Dai}}]{Yang2019AJ....157..111Y}
{Yang}, Y., {Yuan}, H., \& {Dai}, H. 2019, \aj, 157, 111

\bibitem[{{Zahn}(1989)}]{zahn}
{Zahn}, J.~P. 1989, \aap, 220, 112

\bibitem[{Çakırlı {et~al.}(2014)Çakırlı, Ibanoglu, \&
  Sipahi}]{10.1093/mnras/stu958}
Çakırlı, O., Ibanoglu, C., \& Sipahi, E. 2014, Monthly Notices of the Royal
  Astronomical Society, 442, 1560

\end{thebibliography}

\clearpage
\onecolumn
\begin{appendix}
\section{Orbital period tables}
\label{appendixtab}

\begin{table}[h]
\caption{Obtained orbital periods for each data set from our other B+B and O+B overcontact binaries. For each set the BJD from the first data point and the central BJD are given.}
\label{tab:appendix}
    \centering
    \begin{tabular}{c c P{25mm} P{25mm} c c}
        \hline \hline
         Identifier & Catalog & First data point \newline [BJD - 2400000] & Central BJD \newline [BJD - 2400000] & $P$ [s] & $P$ [d] \\
         \hline
                    & ASAS     & 51980 & 53575  & $ 77470.17 \pm 0.07$ & $0.8966455 \pm 0.0000008$ \\
                    & CoRoT    & 54413 & 54449  & $ 77470.30 \pm 0.17$ & $0.896647 \pm 0.000002$ \\
            GU~Mon  & CoRoT    & 54488 & 54508  & $ 77470.4 \pm 0.4 $  & $0.896648 \pm 0.000004$\\
                    & TESS (yr 1)   & 58468 & 58479  & $ 77469.6 \pm 1.8 $  & $0.89664 \pm 0.00002$\\
                    & TESS (yr 3)  & 59202 & 59215  & $ 77469.4 \pm 1.0 $  & $0.89664 \pm 0.000011$\\
        \hline 
                    & DASCH     & 11144 & 14606 & $ 143526.7 \pm 0.4  $      & $1.661189  \pm 0.000005 $  \\
                    & DASCH     & 18069 & 19361 & $ 143506.9 \pm 0.5  $      & $1.660960  \pm 0.000006 $  \\
                    & DASCH     & 20655 & 21737 & $ 143465.6 \pm 0.9  $      & $1.660482  \pm 0.000010 $  \\
                    & DASCH     & 22822 & 25413 & $ 143427.7 \pm 0.4  $      & $1.660043  \pm 0.000004 $  \\
                    & DASCH     & 28004 & 28702 & $ 143415.1 \pm 1.6  $      & $1.659897  \pm 0.000018 $  \\
                    & DASCH     & 29403 & 29975 & $ 143404.1 \pm 1.5  $      & $1.659769  \pm 0.000017 $  \\
                    & DASCH     & 30546 & 31210 & $ 143389.5 \pm 1.4  $      & $1.659601  \pm 0.000016 $  \\
            SV~Cen  & DASCH     & 31876 & 33389 & $ 143379.8 \pm 0.7  $      & $1.659489  \pm 0.000009 $  \\
                    & DASCH     & 40986 & 44380 & $ 143291.3 \pm 0.4  $      & $1.658464  \pm 0.000004 $  \\
                    & ASAS      & 51920 & 53484 & $ 143213.1 \pm 0.3  $      & $1.657559  \pm 0.000004 $  \\
                    & OMC       & 52790 & 53874 & $ 143210.6 \pm 0.4  $      & $1.657531  \pm 0.000004 $  \\
                    & ASAS      & 53358 & 54208 & $ 143209.1 \pm 0.8  $      & $1.657513  \pm 0.000009 $  \\
                    & OMC       & 57596 & 58696 & $ 143194.325 \pm 0.16 $    & $1.657342  \pm 0.000002 $   \\
                    & TESS (yr 1)   & 58569 & 58597 & $ 143205 \pm 4 $          & $1.65748   \pm 0.00005$    \\
                    & TESS (yr 3)   & 59308 & 59334 & $ 143191 \pm 13 $ & $1.65730    \pm 0.00015 $     \\
                    & TESS (yr 5)   & 60041 & 60054 & $ (14322 \pm 5) \cdot 10 $ & $1.6576    \pm 0.0006 $     \\
        \hline 
                    & DASCH     & 11151 & 14794 & $ 129177.3   \pm 0.3  $     & $1.495108    \pm  0.000003  $\\
                    & DASCH     & 18438 & 20085 & $ 129178.0   \pm 0.3  $     & $1.495116    \pm  0.000003  $\\
                    & DASCH     & 21732 & 25416 & $ 129178.1   \pm 0.1  $     & $1.495116    \pm  0.000001  $\\
                    & DASCH     & 29101 & 29936 & $ 129175.2   \pm 1.0  $     & $1.495084    \pm  0.000011  $\\
                    & DASCH     & 30772 & 31518 & $ 129177.4   \pm 1.3  $     & $1.495109    \pm  0.000015  $\\
            V606~Cen& DASCH     & 32264 & 40019 & $ 129175.60  \pm 0.10  $    & $1.4950879   \pm  0.0000012 $\\
                    & ASAS      & 51878 & 53483 & $ 129176.3   \pm 0.2  $     & $1.495096    \pm  0.000002  $\\
                    & OMC       & 52809 & 54826 & $ 129176.58  \pm 0.08  $    & $1.4950993   \pm  0.0000009 $\\
                    & ASAS      & 53357 & 54221 & $ 129176.3   \pm 0.4  $     & $1.495096    \pm  0.000005  $\\
                    & OMC       & 56842 & 58294 & $ 129175.21  \pm 0.10  $    & $1.4950835   \pm  0.0000012 $\\
                    & TESS (yr 1)   & 58601 & 58612 & $ 129207     \pm 14 $       & $1.4954      \pm  0.0002    $\\
                    & TESS (yr 3)   & 59334 & 59347 & $ (12920  \pm 3) \cdot 10 $ & $1.4954      \pm  0.0004    $\\
                    & TESS (yr 5)   & 60041 & 60054 & $ 129170     \pm 3  $   & $1.49502     \pm  0.00003   $\\
        \hline
                    & DASCH    & 11198 & 16176 & $ 65826.50  \pm 0.07 $  & $0.7618808   \pm 0.0000008  $\\
                    & DASCH    & 21156 & 25659 & $ 65826.09  \pm 0.08 $  & $0.7618760   \pm 0.0000009  $\\
            V701~Sco& DASCH    & 30162 & 38967 & $ 65825.89  \pm 0.08 $  & $0.7618737   \pm 0.0000009  $\\
                    & ASAS     & 51938 & 53521 & $ 65825.47  \pm 0.02 $  & $0.7618689   \pm 0.0000002  $\\
                    & TESS (yr 1)  & 58629 & 58640 & $ 65825.7  \pm 1.1 $    & $0.761871    \pm 0.000013   $\\
                    & TESS (yr 3)  & 59362 & 59375 & $ 65825.9  \pm 0.4 $    & $0.761874    \pm 0.000005   $\\
        \hline 
                    & OMC      & 52654 & 55105 & $ 121876.38 \pm 0.10$   & $ 1.41060627  \pm 0.0000012 $\\
            V745~Cas& OMC      & 57556 & 58396 & $ 121876.8  \pm 0.3 $   & $ 1.41061119  \pm 0.000004  $\\
                    & TESS (yr 2)  & 58765 & 58873 & $ 121879.3    \pm 0.5   $   & $ 1.410640  \pm 0.000006   $\\
        \hline
                    & OGLE     & 55260 & 55987 & $ 98597.2 \pm 0.6$       & $1.141172 \pm 0.000007 $ \\
            VFTS~066& TESS (yr 1)  & 58327 & 58504 & $ 98599 \pm 6$       & $1.14119 \pm 0.00007 $ \\
                    & TESS (yr 3)  & 59036 & 59212 & $ 98588 \pm 4$       & $1.12106 \pm 0.00004 $ \\
        \hline
    \end{tabular}
\end{table}

\clearpage
\twocolumn
\section{Orbital period plots}
\label{appendixfig}
\begin{figure}[ht]
    \centering
    \includegraphics[width=0.47\textwidth]{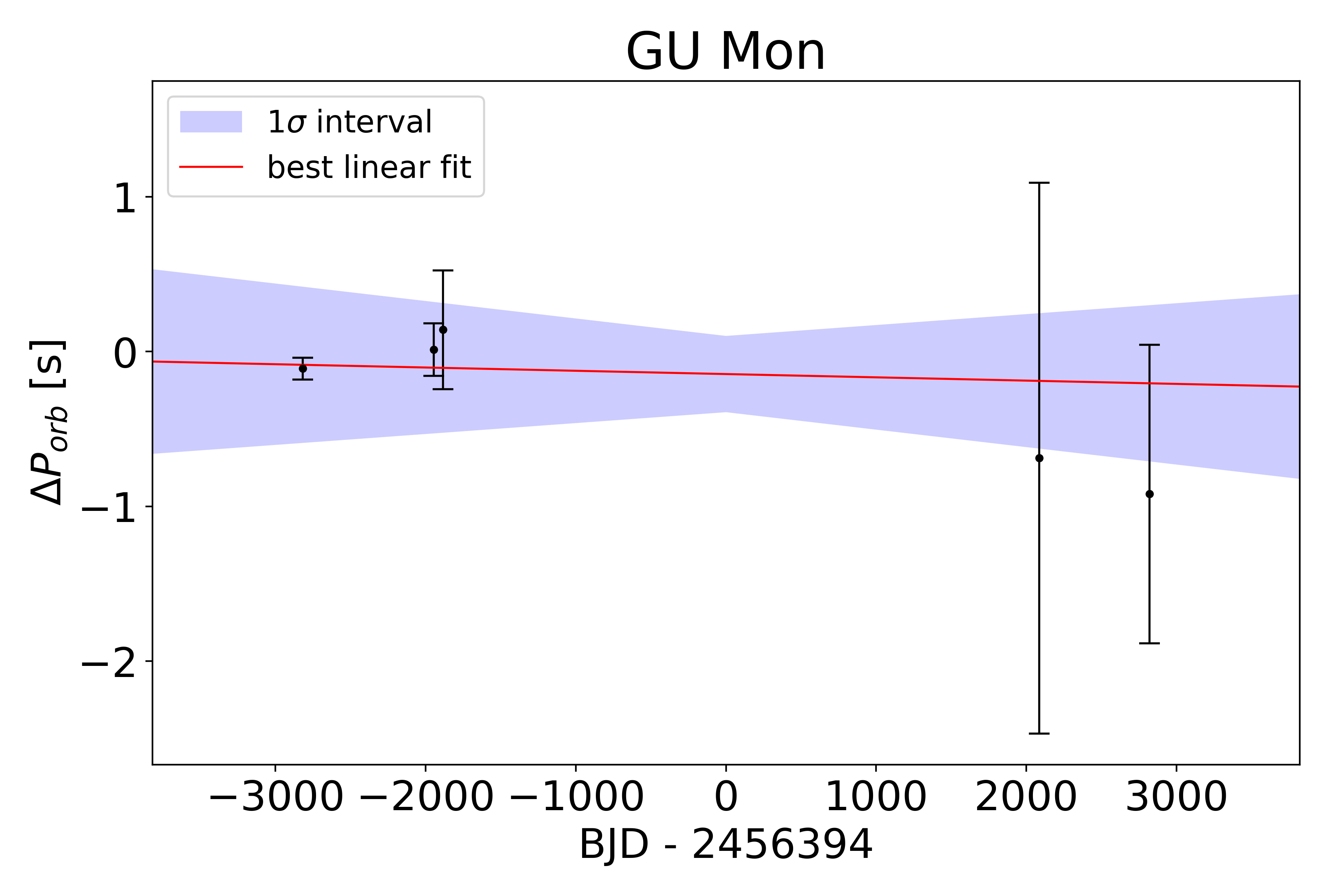}
    \caption{Similar to Fig. \ref{fig:cttau} but for GU~Mon.}
    \label{fig:gumon}
\end{figure}

\begin{figure}[ht]
    \centering
    \includegraphics[width=0.47\textwidth]{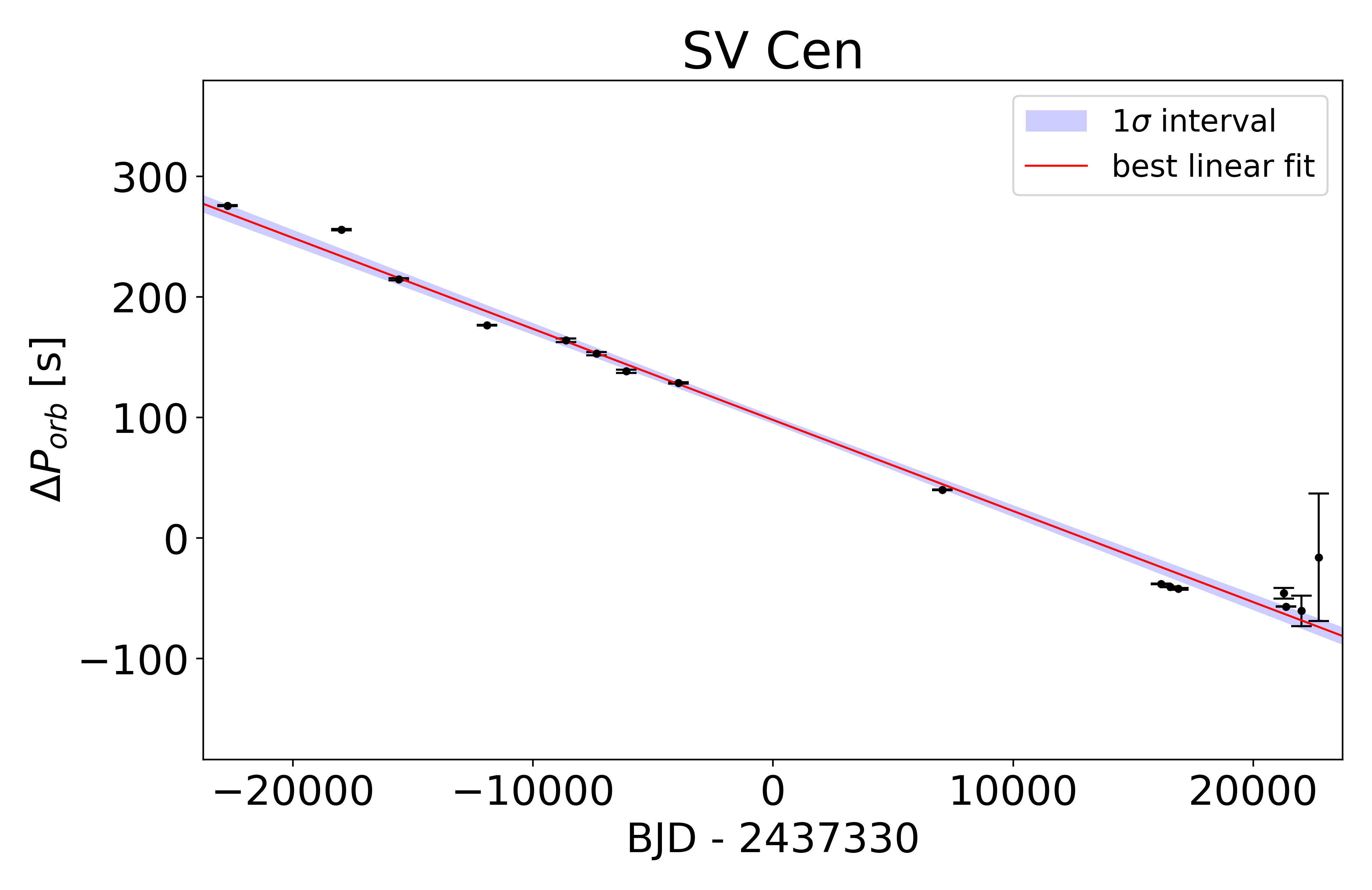}
    \caption{Similar to Fig. \ref{fig:cttau} but for SV~Cen.}
    \label{fig:svcen}
\end{figure}

\begin{figure}[ht]
    \centering
    \includegraphics[width=0.47\textwidth]{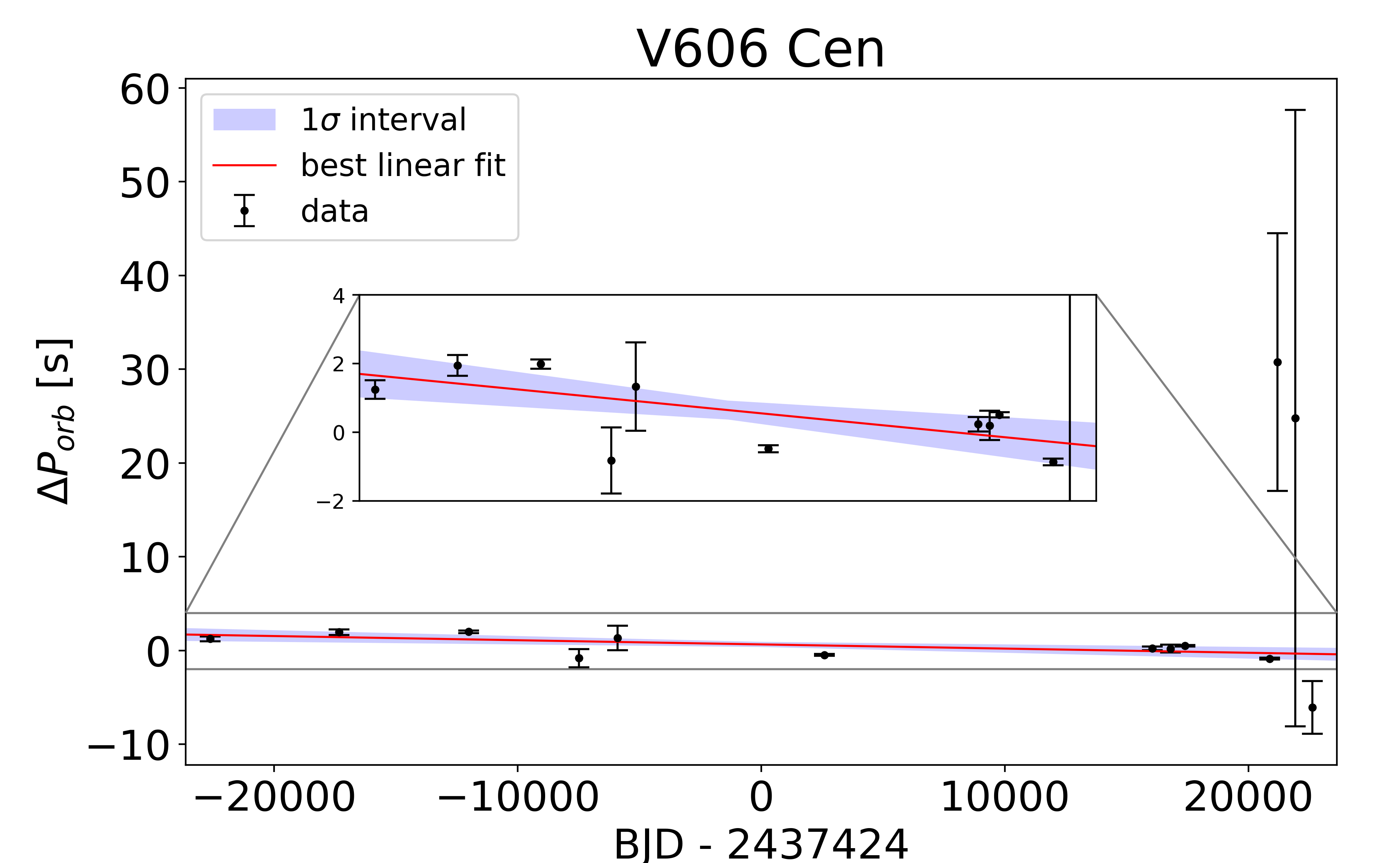}
    \caption{Similar to Fig. \ref{fig:cttau} but for V606~Cen.}
    \label{fig:v606cen}
\end{figure}

\begin{figure}[ht]
    \centering
    \includegraphics[width=0.47\textwidth]{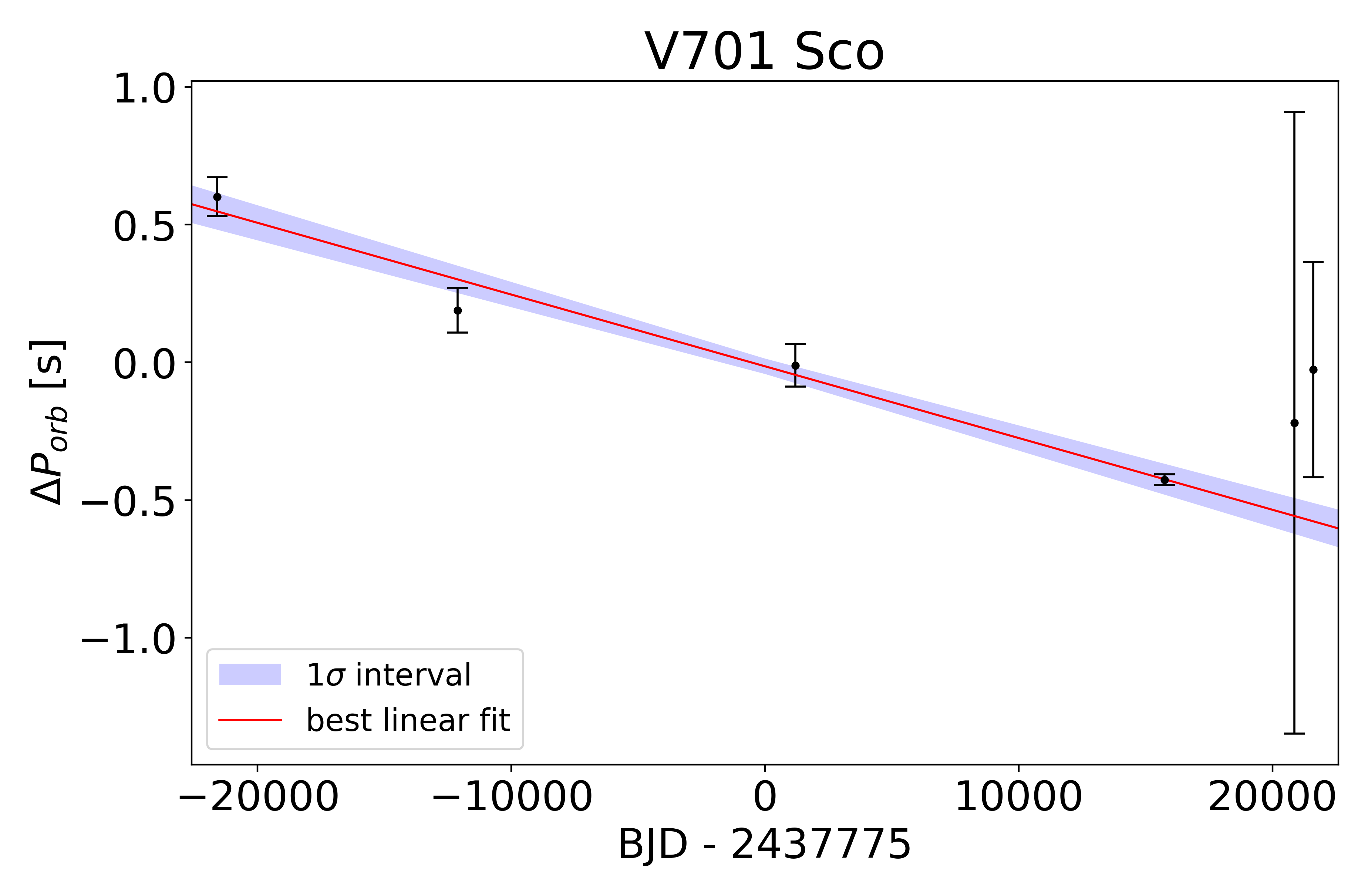}
    \caption{Similar to Fig. \ref{fig:cttau} but for V701~Sco.}
    \label{fig:v701sco}
\end{figure}

\begin{figure}[ht]
    \centering
    \includegraphics[width=0.47\textwidth]{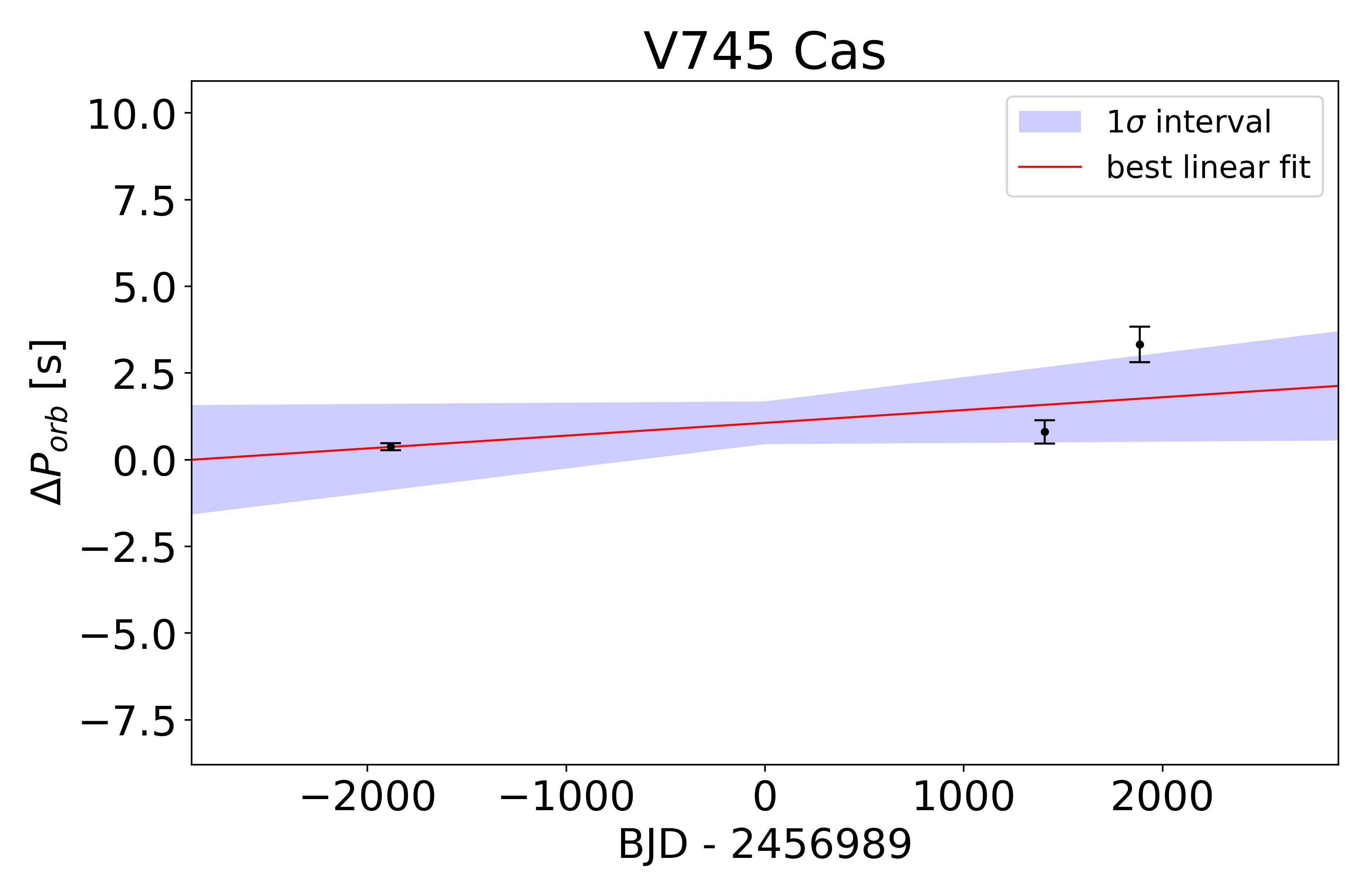}
    \caption{Similar to Fig. \ref{fig:cttau} but for V745~Cas.}
    \label{fig:v745cas}
\end{figure}

\begin{figure}[ht]
    \centering
    \includegraphics[width=0.47\textwidth]{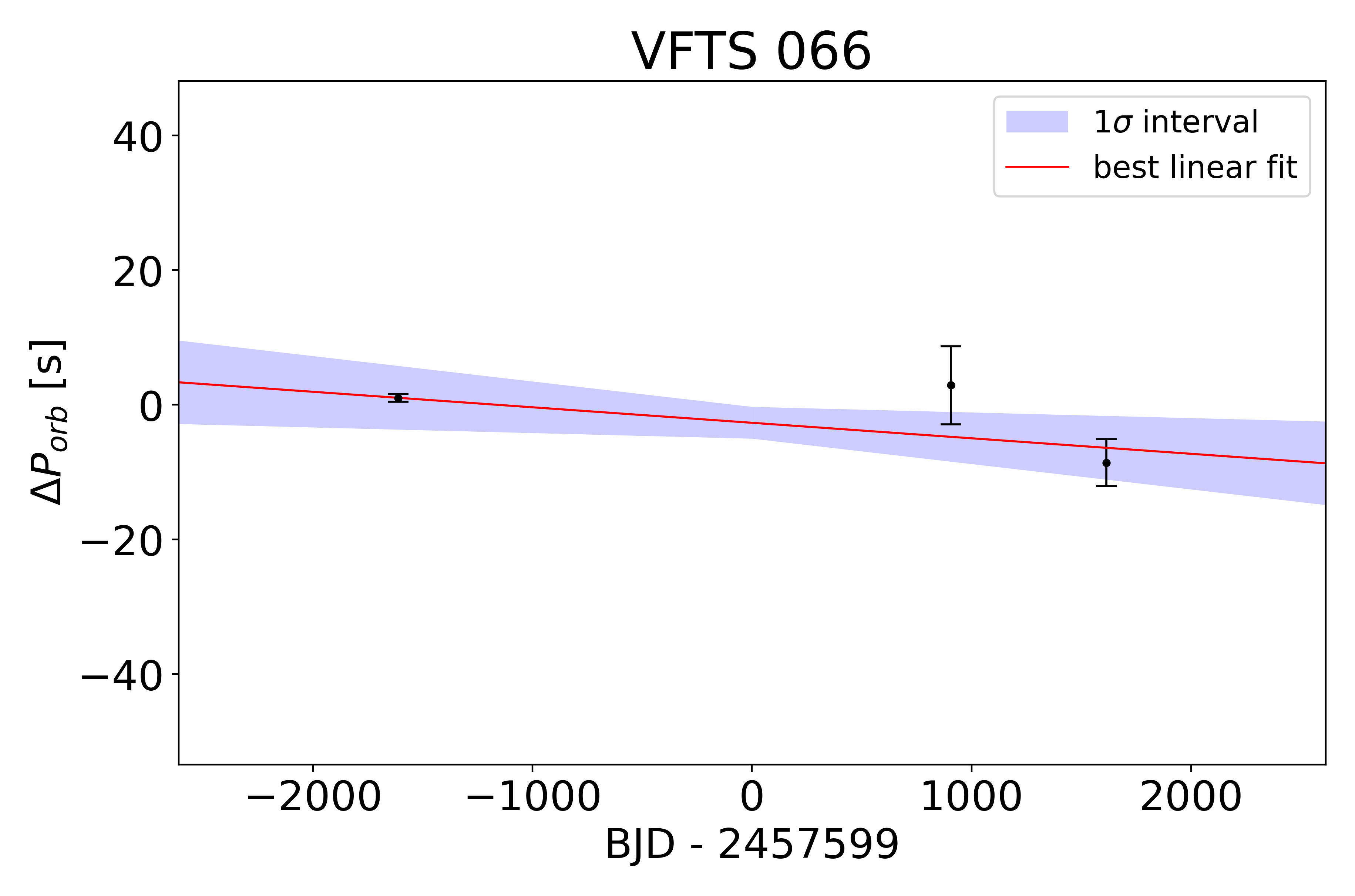}
    \caption{Similar to Fig. \ref{fig:cttau} but for VFTS~066.}
    \label{fig:vfts066}
\end{figure}

\end{appendix}
\end{document}